\documentclass[11pt,preprint]{aastex}

\received{2002 April 25}
\begin{document}

\shortauthors{Welty et al.}
\shorttitle{IV and HV Gas toward $\zeta$ Ori}

\title{Intermediate- and High-Velocity Ionized Gas toward $\zeta$ Orionis\footnotemark\altaffilmark{1}}

\author{Daniel E. Welty}
\affil{University of Chicago, Astronomy and Astrophysics Center, 5640 S. Ellis Ave., Chicago, IL 60637}
\email{welty@oddjob.uchicago.edu}

\author{Edward B. Jenkins}
\affil{Princeton University Observatory, Princeton, NJ 08544}
\email{ebj@astro.princeton.edu}

\author{John C. Raymond}
\affil{Harvard-Smithsonian Center for Astrophysics, 60 Garden St., Cambridge, MA 02138}
\email{jraymond@cfa.harvard.edu}

\author{Christoforos Mallouris}
\affil{University of Chicago, Astronomy and Astrophysics Center, 5640 S. Ellis Ave., Chicago, IL 60637}
\email{chrism@oddjob.uchicago.edu}

\and

\author{Donald G. York\altaffilmark{2}}
\affil{University of Chicago, Astronomy and Astrophysics Center, 5640 S. Ellis Ave., Chicago, IL 60637}
\email{don@oddjob.uchicago.edu}

\altaffiltext{1}{Based in part on observations with the NASA/ESA {\it Hubble Space Telescope}, obtained at the Space Telescope Science Institute, which is operated by the Association of Universities for Research in Astronomy, Inc., under NASA contract NAS5-26555.}

\altaffiltext{2}{also, Enrico Fermi Institute}

\begin{abstract}

We combine near-UV spectra obtained with the {\it Hubble Space Telescope} GHRS echelle with far-UV spectra obtained with IMAPS and {\it Copernicus} to study the abundances and physical conditions in the predominantly ionized gas seen at high velocity ($-$105 km s$^{-1}$ $\la$ $v_{\odot}$ $\la$ $-$65 km s$^{-1}$) and at intermediate velocity ($-$60 km s$^{-1}$ $\la$ $v_{\odot}$ $\la$ $-$10 km s$^{-1}$) along the line of sight to the star $\zeta$ Ori.
We have high resolution (FWHM $\sim$ 3.3--4.5 km s$^{-1}$) and/or high S/N spectra for at least two significant ions of C, N, Al, Si, S, and Fe --- enabling accurate estimates for both the total $N$(\ion{H}{2}) and the elemental depletions. 
C, N, and S have essentially solar relative abundances; Al, Si, and Fe appear to be depleted by about 0.8, 0.3--0.4, and 0.95 dex, respectively, relative to C, N, and S.
While various ion ratios would be consistent with collisional ionization equilibrium (CIE) at temperatures of 25,000--80,000 K, the widths of individual high-velocity absorption components indicate that $T$ $\sim$ 9000$\pm$2000 K --- so the gas is not in CIE.
Analysis of the \ion{C}{2} fine-structure excitation equilibrium, at that temperature, yields estimates for the densities ($n_e$ $\sim$ $n_{\rm H}$ $\sim$ 0.1--0.2 cm$^{-3}$), thermal pressures ($2 n_{\rm H} T$ $\sim$ 2000--4000 cm$^{-3}$K), and thicknesses (0.5--2.7 pc) characterizing the individual clouds.
We compare the abundances and physical properties derived for these clouds with those found for gas at similar velocities toward 23 Ori and $\tau$ CMa, and also with several different models for shocked gas.
While the shock models can reproduce some features of the observed line profiles and some of the observed ion ratios, there are also significant differences between the models and the data.
The measured depletions suggest that roughly 10\% of the Al, Si, and Fe originally locked in dust in the pre-shock medium may have been returned to the gas phase, consistent with recent predictions for the destruction of silicate dust in a 100 km s$^{-1}$ shock.
The observed near-solar gas phase abundance of carbon, however, appears to be inconsistent with the predicted longer time scales for the destruction of graphite grains.
\end{abstract} 


\section{Introduction}
\label{sec-intro}

Interstellar gas found at high and intermediate velocities in the Galactic plane near associations of young, massive stars presumably reflects the combined effects of stellar winds and supernovae arising in those associations.
Observations of UV absorption lines can reveal the kinematics, elemental abundances (and depletions), ionization state, and physical properties (density, temperature, etc.) characterizing such high- and intermediate-velocity gas --- and can thus provide significant empirical tests for models of (1) the interaction of supernovae with the surrounding interstellar medium, (2) shocks and the non-equilibrium evolution of shocked gas, and (3) the survival of dust grains in such environments.

UV spectra obtained with {\it Copernicus} revealed absorption from various singly and doubly ionized species at both high ($-$100 to $-$80 km s$^{-1}$) and intermediate ($-$80 to $-$15 km s$^{-1}$) heliocentric velocities toward a number of stars in Orion (Cohn \& York 1977; Cowie, Songaila, \& York 1979).
Cowie et al. ascribed the widespread high-velocity (HV) gas to a radiative shock produced by one or more recent supernovae and called it ``Orion's Cloak''.
Subsequent UV spectra of additional stars in Orion obtained with {\it IUE} provided further information on the distribution of the HV and intermediate-velocity (IV) gas throughout that region (Shore 1982; Sonneborn, Shore, \& Brown 1988).
More recently, higher resolution UV spectra of 23 Ori obtained with the {\it Hubble Space Telescope} Goddard High Resolution Spectrograph ({\it HST}/GHRS) have enabled more precise characterization of the HV and IV gas in that line of sight (Trapero et al. 1996; Welty et al. 1999).
Toward 23 Ori, the HV gas is essentially fully ionized, and is characterized by a density $n_e$ $\sim$ $n_{\rm H}$ $\sim$ 0.4--0.5 cm$^{-3}$, a temperature $T$ $\sim$ 8000 K, and a thermal pressure 2$n_{\rm H}T$ $\sim$ 5--10 $\times$ 10$^3$ cm$^{-3}$K.
As the ion ratios $N$(\ion{X}{3})/$N$(\ion{X}{2}) for aluminum and silicon would be consistent with collisional ionization equilibrium (CIE) at much higher temperatures --- of order 25000 K --- the HV gas is clearly not in CIE.
No data (or only upper limits) were available, however, for a number of potentially significant ions (e.g., \ion{C}{3}, \ion{N}{3}, \ion{S}{3}, \ion{Fe}{2}, \ion{Fe}{3}), and some of the spectra were obtained at only moderate spectral resolution.

In this paper, we discuss the predominantly ionized HV and IV gas toward a second Orion star ($\zeta$ Ori), using near-UV spectra obtained with GHRS and far-UV spectra obtained with the Interstellar Medium Absorption Profile Spectrograph (IMAPS) and {\it Copernicus}.
In Sections~\ref{sec-obsred} and~\ref{sec-prof}, we describe the spectra and our analysis of the observed absorption-line profiles.
In Section~\ref{sec-res}, we discuss the abundances and physical properties derived for the individual components in the HV and IV gas.
The unprecedented combination of high spectral resolution (FWHM $\sim$ 3.3--4.5 km s$^{-1}$), high signal-to-noise ratio (S/N), and broad spectral coverage has enabled the determination of accurate abundances for both singly and doubly ionized C, N, Al, Si, S, and Fe --- which then provide strong constraints on the overall $N$(H), the depletions, and the ionization in the HV and IV gas. 
Detailed multicomponent fits to the high-resolution line profiles and analyses of the \ion{C}{2} fine-structure excitation equilibrium yield values for the temperature and density of the gas, respectively.
In Section~\ref{sec-disc}, we compare the elemental abundances and derived physical properties with those found for gas observed at similar velocities in several other lines of sight, we compare the observed line profiles and ion ratios [$N$(\ion{X}{3})/$N$(\ion{X}{2}), $N$(\ion{X}{4})/$N$(\ion{X}{2})] with those predicted by various models for ionized gas (including newly-computed models for shocked gas), and we discuss the observed depletions in relation to recent models for dust processing in shocks.
In an appendix, we present similar (though more limited) data for the high-velocity gas toward $\tau$ CMa.

\section{Observations and Data Reduction}
\label{sec-obsred}

\subsection{$\zeta$ Ori Line of Sight}
\label{sec-los}

The O9.5 Ib star $\zeta$ Ori A [V = 1.88; $E(B-V)$ = 0.06; $v_{\rm rad}$ = 18.1 km s$^{-1}$; $v$sin$i$ = 135 km s$^{-1}$] is located at ($l$, $b$) = (206$\fdg$45, $-$16$\fdg$59), at the southeast end of the belt of Orion.
(Unless otherwise specified, we will use heliocentric velocities; the zero point for velocities with respect to the Local Standard of Rest is at v$_{\odot}$ $\sim$ +17.6 km s$^{-1}$.)
The parallax 3.99$\pm$0.79 mas measured by {\it Hipparcos} implies a distance of 251$^{+62}_{-42}$ pc (ESA 1997), consistent with, but perhaps somewhat smaller than the value 340$\pm$85 pc inferred from the spectral type and apparent magnitude (using the absolute magnitudes of Blaauw 1963).
Brown, Walter, \& Blaauw (1998) have suggested that Orion OB association subgroup 1b (of which $\zeta$ Ori is a member) lies at an average  distance of about 430 pc, however, based on {\it Hipparcos} parallaxes for a number of Orion stars.
The line of sight to $\zeta$ Ori has a total neutral hydrogen column density log[$N$(\ion{H}{1} + 2H$_2$)] = 20.43 cm$^{-2}$ (Bohlin, Savage, \& Drake 1978; Shull \& van Steenberg 1985; Diplas \& Savage 1994), with only a small fraction in molecular form (log[$N$(H$_2$)] $\sim$ 15.85 cm$^{-2}$; Jenkins \& Peimbert 1996).
$\zeta$ Ori has been observed in connection with a number of surveys of optical and UV interstellar absorption-lines (e.g., Hobbs 1969; Marschall \& Hobbs 1972; van Steenberg \& Shull 1988; Welty, Hobbs, \& Kulkarni 1994; Welty, Morton, \& Hobbs 1996), some properties of the primarily neutral gas at low velocities have been discussed by Drake \& Pottasch (1977) and by Jenkins \& Peimbert (1996), and the high-velocity ionized gas associated with Orion's Cloak has been described by Cowie et al. (1979).

\subsection{GHRS Spectra}
\label{sec-obsghrs}

We observed $\zeta$ Ori with the {\it HST} GHRS during Cycle 5 (on 1996 Feb 12), as part of GO program 5896 for studying high-velocity ionized gas.
All observations reported here were obtained through the small science aperture, in order to achieve the highest possible spectral resolution.
Eight wavelength regions between 1188 and 1676 \AA\ and six regions between 1740 and 2805 \AA, each covering 6--15 \AA, were observed using the echelle grating at resolutions of 3.3--3.7 km s$^{-1}$ (Table~\ref{tab:ghrs}).
In order to obtain proper sampling and to reduce the effects of photocathode sensitivity variations, all exposures were divided into a number of slightly shifted sub-exposures, using FP-SPLIT = DSFOUR and COMB = 4 (see, e.g., Duncan \& Ebbets 1989).
Approximately 6\% of each exposure was devoted to measuring the inter-order background.
Because the cores of many of the obviously saturated absorption lines at low velocities fell either a few percent above or below the zero level in the extracted spectra received from STScI, the spectra were re-extracted using the STSDAS CALHRS routines, with a cubic polynomial fit to the background (instead of the default constant background).
Even after reprocessing, the cores of some saturated lines still differed slightly (generally less than 1\%) from zero; those spectra were adjusted accordingly.
In addition, some G160M spectra of $\zeta$ Ori (originally acquired under GO program 2584) were obtained from the {\it HST} archive in order to measure absorption lines from species not covered in our spectra.
Nine wavelength regions between 1186 and 1875 \AA, each covering approximately 36--38 \AA, had been observed with grating G160M at intermediate resolutions of 13--20 km s$^{-1}$ (Table~\ref{tab:ghrs}).

Because $\zeta$ Ori is very bright, even these short GHRS exposures (3.5--5.5 min) were expected to yield S/Ns of 100--500 per half resolution element in the extracted spectra --- if the detector fixed-pattern noise (FPN) could be adequately removed.
For the ECH-A and ECH-B spectra, the STSDAS routines POFFSETS, DOPOFF, and SPECALIGN were initially used to determine the detector fixed-pattern features and to align and sum the sub-exposures.
Near most of the absorption features in the spectra, application of those routines did yield S/N nearly as high as those expected from photon statistics.
Near the extreme ends of some of the spectra, and in cases where the fixed-pattern features differed in strength among the sub-exposures, however, those routines did not completely remove the FPN.
An alternative FPN removal procedure based on Fourier transform methods (developed by EBJ) was therefore also used to correct and combine the sub-exposures.
In general, the two procedures yielded very similar combined, corrected spectra.
The STSDAS routines performed better near the cores of the strongest absorption features, while the Fourier technique was more successful in removing FPN features which differed in strength among the sub-exposures (e.g., near the HV absorption for the \ion{Fe}{2} $\lambda$2600 line). 

For the G160M exposures obtained from the archive, the relative offsets among the subexposures in the nominal wavelength scales supplied by STScI were determined by fitting the profiles of interstellar and/or stellar lines in the spectra.
The individual subexposures were shifted according to those relative offsets (which were always less than 13 km s$^{-1}$), sinc interpolated (Spitzer \& Fitzpatrick 1993) onto a roughly twice-oversampled standard wavelength grid, and summed.
Obvious deviant pixels in the individual subexposures (due to detector pattern features, bad pixels, radiation hits, etc.) were given zero weight in the sums; no other explicit detector pattern noise identification and removal was performed, however.
Adjacent pixels were then summed to yield approximately optimal sampling (i.e., two points per resolution element) in the final spectra.

Portions of the spectra near interstellar absorption lines of interest (usually including 150--250 points) were normalized by fitting the continuum regions with Legendre polynomials of orders typically 4--9.
The continuum signal-to-noise ratios achieved for the echelle spectra range from 100--550 (Table~\ref{tab:ghrs}); the corresponding equivalent width uncertainties (2$\sigma$) for weak, unresolved absorption features range from 0.2--1.0 m\AA.
For the G160M data, the S/N are typically between 60 and 350, yielding 2$\sigma$ equivalent width upper limits or uncertainties of 1-5 m\AA.
For some lines with strong absorption over a wide velocity range (e.g., \ion{C}{2} $\lambda$1334, \ion{Si}{3} $\lambda$1206, and the stronger \ion{Si}{2} lines), the adopted continuum is less well determined.
The increased uncertainty in continuum placement affects primarily the absorption at low and intermediate velocities.
The quoted equivalent width errors include contributions from both photon noise (Jenkins et al. 1973) and continuum placement (Sembach \& Savage 1992), added in quadrature.

Some of the normalized GHRS spectra, together with similar spectra from IMAPS (\S~\ref{sec-obsimaps}) and {\it Copernicus} (\S~\ref{sec-obsoao}), are shown in Figure~1.
This figure includes the absorption at both low and high velocities;
the high- and intermediate-velocity absorption features (as defined in \S~\ref{sec-prof}) will be shown in more detail in Sections~\ref{sec-dsmod1} and \ref{sec-dsmod2}, where they are compared with those predicted by several shock models.
The equivalent widths (or limits) for the high-velocity absorption due to various atomic and ionic species found in the UV spectra are listed in Table~\ref{tab:ew}.
The intermediate-velocity absorption features are often blended with the strong absorption due to primarily neutral gas at lower velocities. 
The source of each observed line (GHRS echelle, GHRS G160M, IMAPS, {\it Copernicus}) is noted following the wavelength.
Dominant contributors to blended lines are marked with a ``B'' following the equivalent width; minor contributors to blended lines are included in the table, but are referred to the dominant contributors for the total equivalent widths.

\subsection{IMAPS Spectra}
\label{sec-obsimaps}

The far-UV spectrum of $\zeta$ Ori between about 950 and 1150 \AA\ was observed with IMAPS during that instrument's first orbital flight on the ORFEUS-SPAS platform, carried by Space Shuttle flight STS-51 in 1993 September.
Jenkins et al. (1996) have given a very detailed description of the IMAPS payload, its performance during that orbital flight, and the data reduction procedures employed to obtain wavelength calibrated, extracted spectra from the numerous two-dimensional partial echelle images typically acquired for each target.
Sixty-three CCD frames, each covering about 1/4 the free spectral range of the echelle, were obtained in 2412 seconds of total integration on $\zeta$ Ori.
These images were corrected for electronic interference effects, dark current, non-uniformities in photocathode response, and geometric distortion as described by Jenkins et al., and the individual frames were aligned and co-added.
The one-dimensional spectra were obtained via an optimal extraction routine which took into account the contamination from the wings of adjacent orders.

The resolution achieved in the IMAPS spectra of $\zeta$ Ori is approximately 4.5 km s$^{-1}$ (FWHM; Jenkins \& Peimbert 1996).
We therefore smoothed the original oversampled spectra, by adding adjacent data points, to obtain (roughly) two pixels per resolution element.
The continua near lines of interest were fitted with low-order Legendre polynomials.
The S/Ns measured in the IMAPS spectra are generally between about 30 and 80 for wavelengths between 1000 and 1150 \AA, but can be significantly smaller below 1000 \AA.
Background levels are typically accurate to within about 5\% (Sofia \& Jenkins 1998); corrections could be made in cases where there are nearby saturated low-velocity absorption features.
The velocity zero point was established by using the various telluric \ion{O}{1}* and \ion{O}{1}** features present in the spectrum.

Jenkins \& Peimbert (1996) have discussed the numerous H$_2$ absorption features arising in the low-velocity gas toward $\zeta$ Ori.
In this study of the ionized gas, we focus on the absorption lines of \ion{C}{2}, \ion{C}{3}, \ion{N}{2}, \ion{N}{3}, \ion{S}{3}, \ion{S}{4}, and \ion{Fe}{3} found in the IMAPS spectra.
Several of the normalized IMAPS spectra are shown in Figure~1 and the equivalent widths measured from those spectra are listed in Table~\ref{tab:ew}. 
The quoted uncertainties include contributions from photon noise, continuum placement, and background level --- added in quadrature.

\subsection{{\it Copernicus} Spectra}
\label{sec-obsoao}

\begin{figure}
\figurenum{1a}
\epsscale{0.8}
\plottwo{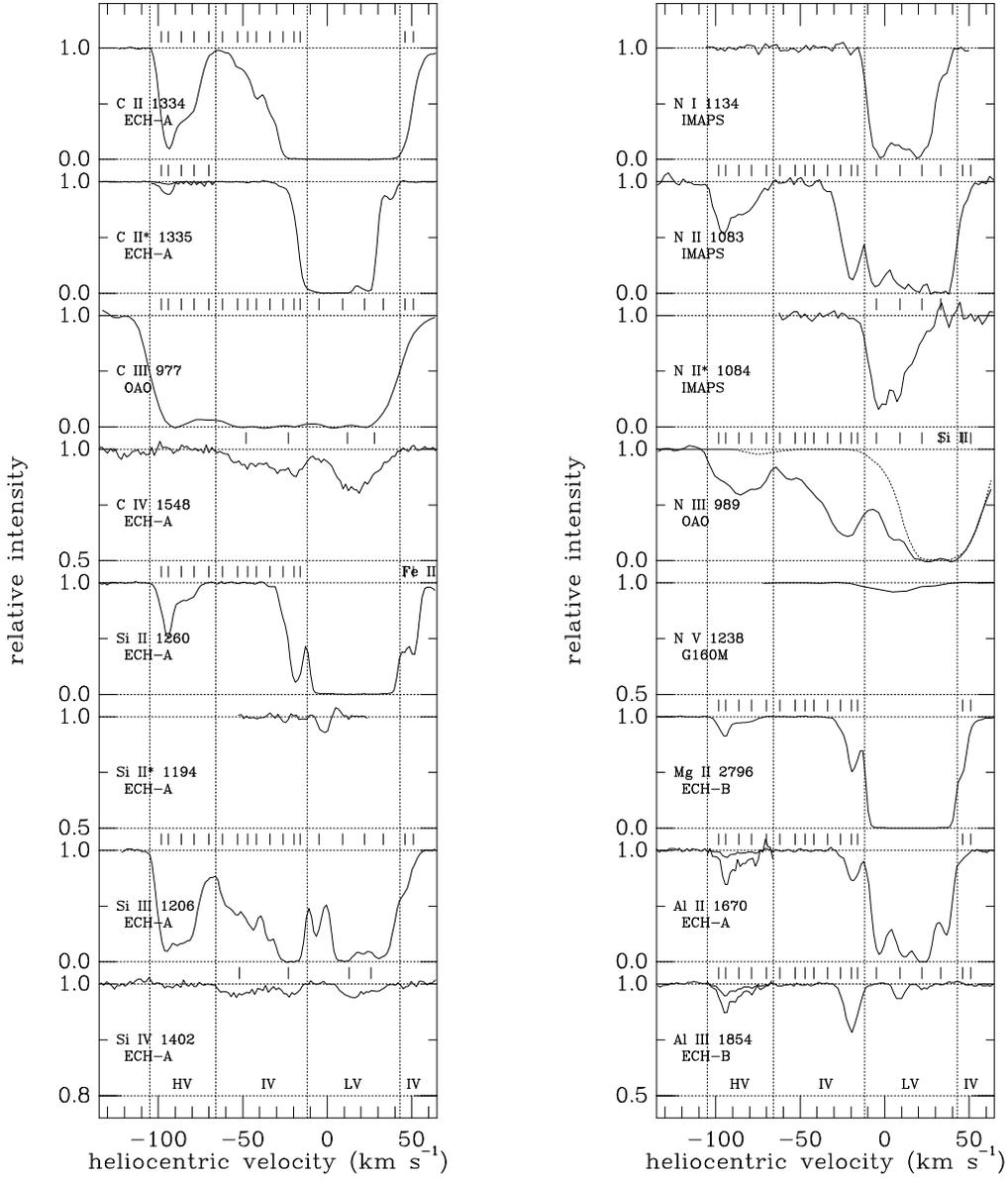}{f1b.eps}
\caption{Spectra of UV absorption lines observed toward $\zeta$ Ori.
The source of each profile is noted below the line identification.
Tick marks denote the components derived from fits to the spectra.
The vertical dotted lines separate the high-velocity (HV), intermediate-velocity (IV), and low-velocity (LV) component groups.
Absorption due to other species is noted above some of the spectra; the dotted line in the N III spectrum shows the absorption due to Si II, and the dotted line in the S III spectrum shows the absorption due to Si II.
Note the expanded vertical scale for some of the weaker lines; where two lines are plotted for the HV gas, the lower has been expanded by a factor of 5 (C II*, Al II, S II, S III, Fe II) or of 2.5 (Al III).}
\label{fig:spec1}
\end{figure}

Relatively high S/N far-UV spectra of $\zeta$ Ori had also been obtained with {\it Copernicus} by Cowie et al. (1979) for their study of the ionized high-velocity gas in the Orion region.
Although of somewhat lower resolution (FWHM $\sim$ 0.045 \AA, corresponding to 13.5 km s$^{-1}$ at 1000 \AA), these spectra were retrieved from the {\it Copernicus} archive\footnotemark\ as a supplement to the IMAPS spectra in cases where the latter were of low S/N (e.g., for \ion{C}{3} and \ion{N}{3}).
\footnotetext{Maintained at http://archive.stsci.edu/copernicus/index.html.  See Rogerson et al. (1973) and subsequent articles for descriptions of the {\it Copernicus} (OAO-3) satellite and spectrometer and its early performance.}
The {\it Copernicus} spectra were also useful for confirming the weak absorption due to \ion{S}{4} (at low velocities) and \ion{Fe}{3} (at high velocities) in the IMAPS spectra.
In general, we chose the group of contemporaneous U1 scans with the widest spectral coverage encompassing each line of interest --- to minimize thermal shifts and sensitivity variations and to reach the continuum on either side of the absorption feature.
The individual scans were co-added using a version of the {\it Copernicus} STACK program; continua were fitted using low-order Legendre polynomials.
Background levels could be established for most lines either via reference to the typically saturated absorption at low velocities or by comparison with the corresponding IMAPS spectra.
The normalized spectra of \ion{C}{3} and \ion{N}{3} are shown in Figure~\ref{fig:spec1} (with source name OAO) and the equivalent widths measured from those spectra are listed in Table~\ref{tab:ew}. 
These spectra exhibit very good agreement with those shown by Cowie et al. (1979) --- which suggests that we have used the same sets of scans for each line.

\section{Profile Analysis}
\label{sec-prof}

The interstellar absorption observed toward $\zeta$ Ori is complex.
Very high resolution (FWHM $\sim$ 0.3-1.3 km s$^{-1}$) optical spectra of \ion{Na}{1}, \ion{Ca}{2}, \ion{Ca}{1}, and \ion{K}{1} reveal at least 20 components at heliocentric velocities between about $-$8 and +37 km s$^{-1}$ (Welty et al. 1994, 1996, 2002; Welty \& Hobbs 2001).
These components can be associated with four low-velocity (LV) component groups, centered at about $-$2, +12, +24, and +37 km s$^{-1}$, that are discernible in the lower resolution GHRS echelle and IMAPS spectra of many neutral and singly ionized species.
Several weak intermediate-velocity (IV) components (defined here as having 
$-$60 km s$^{-1}$ $\la$ $v_{\odot}$ $\la$ $-$10 km s$^{-1}$ or 
45 km s$^{-1}$ $\la$ $v_{\odot}$ $\la$ 95 km s$^{-1}$ --- 
i.e., 27.5 km s$^{-1}$ $\la$ $|v_{\rm LSR}|$ $\la$ 77.5 km s$^{-1}$), 
between $-$32 and $-$14 km s$^{-1}$ and also near +50 km s$^{-1}$, can be seen in the profiles of the stronger lines of some first ions (e.g., \ion{Mg}{2}, \ion{Al}{2}, \ion{Si}{2}).
Still weaker IV components, between $-$60 and $-$40 km s$^{-1}$, may be discerned in the very strong \ion{C}{2} $\lambda$1334 and \ion{Si}{3} $\lambda$1206 lines.
The high-velocity (HV) gas comprising Orion's Cloak appears at $-$105 km s$^{-1}$ $\la$ $v$ $\la$ $-$65 km s$^{-1}$ in the strongest lines of various singly and doubly ionized species; the echelle profiles of those ions suggest that several components are present.
The strongest absorption from most doubly ionized species is found in components near $-$16 and +13 km s$^{-1}$, though \ion{C}{3}, \ion{Si}{3}, and (probably) \ion{N}{3} exhibit absorption over essentially the full range from $-$100 to +45 km s$^{-1}$.
Weak, broad absorption due to the higher ions \ion{C}{4} and \ion{Si}{4} extends from about $-$60 to +40 km s$^{-1}$.
Several components, which may be associated with those seen for the lower ions, appear to be present.
A schematic diagram of the observed absorption is given in Figure~\ref{fig:schem}.

We have used the method of profile fitting in an attempt to discern and characterize the individual components giving rise to the observed absorption profiles (e.g., Welty et al. 1991, 1999).
In this analysis, we attempt to obtain column densities ($N$), Doppler line width parameters ($b$ = $[(2kT/m) + 2v_t^2]^{1/2}$) and heliocentric velocities ($v$) for as many components as are needed (at minimum and consistent with the measured S/N) to reproduce the observed line profiles.
The line widths provide direct limits on the temperature ($T$) and turbulent velocity ($v_t$) characterizing the individual components.
The profile analysis provides a means for estimating the properties of heavily blended lines and enables us to model lower resolution spectra via the detailed structure derived from higher resolution line profiles (e.g., to obtain more accurate total column densities).
We assume that each component may be represented by a (symmetric) Voigt profile (though see the discussion of shock models below), smeared by the appropriate instrumental response function.
The instrumental function is assumed to be Gaussian (GHRS and IMAPS) or triangular ({\it Copernicus}), with the FWHM given in Table~\ref{tab:ghrs} or in the discussions above.
We use the rest wavelengths, oscillator strengths, and damping constants tabulated by Morton (1991), except where more recent work has provided more accurate values (Table~\ref{tab:ew}; see the Appendix in Welty et al. 1999).
The total sightline column densities, summed over all the components found in fitting the line profiles, are consistent with the corresponding values derived via apparent optical depth methods (e.g., Savage \& Sembach 1991; Jenkins 1996).

The component structure for the high- and intermediate-velocity gas, derived primarily from the high-resolution UV profiles of \ion{C}{2}, \ion{N}{2}, \ion{Mg}{2}, \ion{Si}{2}, and \ion{Si}{3}, is given in Table~\ref{tab:h2comp}.
The components are numbered sequentially, with the letters H and I appended to associate them with the high and intermediate velocity component groups, and are marked above the spectra in Figure~1.
For the HV gas, the component parameters are most reliable for component 2H at $-$93 km s$^{-1}$ (the strongest and best defined) and for component 4H at $-$78 km s$^{-1}$ (with $b$ constrained by the right-hand wing).
Two additional components (1H at $-$97 km s$^{-1}$ and 3H at $-$86 km s$^{-1}$) are required to fit the absorption shortward of component 2H and between components 2H and 4H, respectively; another very weak component (5H at about $-$68 km s$^{-1}$) is found for \ion{Si}{3} and \ion{C}{2}.
Slight asymmetries in the strongest IV absorption near $-$16 km s$^{-1}$ (e.g., in the lines of \ion{Fe}{2}, \ion{S}{2}, and \ion{Si}{2}) suggest that at least two components (12I at $-$18 km s$^{-1}$, 13I at $-$14 km s$^{-1}$) are present there.
IV components 6I through 9I (between $-$59 and $-$40 km s$^{-1}$) are visible in the \ion{Si}{3} $\lambda$1206, \ion{N}{3} $\lambda$989, and \ion{C}{2} $\lambda$1334 lines.
The $b$-values in parentheses were determined in trial fits to the line profiles, but then were fixed to facilitate convergence in the final iterative fits.
The component velocities and $b$-values found in fitting the stronger HV and IV absorption features  were then used in fitting the weaker absorption from \ion{Al}{2}, \ion{Al}{3}, \ion{S}{2}, \ion{S}{3}, \ion{Fe}{2}, and \ion{Fe}{3}, and also the lines of \ion{C}{3} and \ion{N}{3} observed at lower resolution with {\it Copernicus}.
At least four components appear to be present for the higher ions \ion{C}{4} and \ion{Si}{4}, which were fitted independently.

The absolute velocities for the various spectra were determined by fitting the low-velocity absorption due to (primarily) neutral gas, using the detailed component structure derived from very high resolution optical spectra to model the lower resolution UV spectra (Welty et al. 1999).
The measured velocity offsets between the GHRS echelle spectra and the optical spectra ranged from about +4 km s$^{-1}$ for the first ECH-A spectra obtained to about $-$3.5 km s$^{-1}$ for the last ECH-B spectra.
The corresponding offsets for the IMAPS and {\it Copernicus} spectra are about 2 km s$^{-1}$ and 7 km s$^{-1}$, respectively.

\section{Abundances, Depletions, and Physical Conditions}
\label{sec-res}

Table~\ref{tab:h2abund} gives total column densities (or limits) for various neutral and ionized species in the HV and IV gas.
The values for the heavy element species are based on the profile fits and/or the equivalent widths; the estimates for $N$(\ion{H}{1}) and $N$(\ion{H}{2}) are discussed below.
Figure~\ref{fig:abund} shows the HV and IV column densities as functions of ionization potential. 
By combining near-UV spectra from GHRS with far-UV spectra from IMAPS and {\it Copernicus}, we have HV and IV column densities for both singly and doubly ionized states of C, N, Al, Si, S, and Fe.
We also have significantly restrictive limits for the corresponding neutral species and (for the HV gas) for the triply ionized states of C, Si, and S; \ion{N}{4} is also unlikely to be significant in the HV gas, given the limits on \ion{C}{4}, \ion{Si}{4} and \ion{S}{4}.

The elements nitrogen and sulfur typically have nearly solar abundances (within about 20\%) in diffuse Galactic clouds; carbon is generally very slightly depleted (by a factor of about 2.5); Al, Si, and Fe are usually more severely depleted (factors of 4--300) (e.g., Savage \& Sembach 1996; Fitzpatrick 1996; Welty et al. 1999; and references therein).
We use the meteoritic (C1 carbonaceous chondritic) reference abundances of Anders \& Grevesse (1989) for all elements except C, N, and O, for which we use solar photospheric values from Grevesse \& Noels (1993).\footnotemark\
The total hydrogen column densities $N$(\ion{H}{1})+$N$(\ion{H}{2}) for the HV and IV gas may therefore be estimated from the sums $N$(\ion{N}{2})+$N$(\ion{N}{3}), $N$(\ion{S}{2})+$N$(\ion{S}{3}), and (perhaps) $N$(\ion{C}{2})+$N$(\ion{C}{3}); the level of depletion may be gauged via the corresponding sums for Al, Si, and Fe (assuming negligible \ion{Al}{4} and \ion{Fe}{4}).
The column density limits for \ion{O}{1}, the dominant ionization state of oxygen in neutral (\ion{H}{1}) gas, may be used to set corresponding limits on the column density of \ion{H}{1} associated with each component group.
\footnotetext{Use of the revised solar photospheric abundances for the few elements reported by Holweger (2001) would have a minor impact on the results discussed in this paper.}

Table~\ref{tab:h2prop} summarizes some of the physical properties of the individual HV and IV components toward $\zeta$ Ori.
Using estimates for the temperature derived from the component line widths (\S~\ref{sec-reshv}), analysis of the \ion{C}{2} fine-structure excitation equilibrium yields estimates for the electron density ($n_e$), which is approximately equal to the local hydrogen density ($n_{\rm H}$) for these predominantly ionized components.
(If helium, with first ionization potential 24.6 eV, were entirely singly ionized, then $n_e$ would be roughly 10\% higher than $n_{\rm H}$.)
Dividing the total hydrogen column densities by $n_{\rm H}$ then yields the thicknesses ($r$) of the individual components.
Table~\ref{tab:hvgprop} compares various ion ratios and derived physical properties (averaged over all components) for the HV and IV gas seen toward $\zeta$ Ori, 23 Ori (Welty et al. 1999), and $\tau$ CMa (appendix) with predictions from shock models (discussed below).

\subsection{High-Velocity Gas ($-$105 km s$^{-1}$ $\la$ $v_{\odot}$ $\la$ $-$65 km s$^{-1}$)}
\label{sec-reshv}

The column densities of \ion{C}{2} + \ion{C}{3}, \ion{N}{2} + \ion{N}{3}, and \ion{S}{2} + \ion{S}{3} suggest that C, N, and S are present in nearly solar proportions (within $\pm$25\%) in the HV gas --- and thus that carbon (as well as nitrogen and sulfur) may be at most only very mildly depleted.
Under the assumption that nitrogen and sulfur are undepleted, we estimate log[$N$(\ion{H}{1}) + $N$(\ion{H}{2})] (cm$^{-2}$) $\sim$ 17.88$\pm$0.06 for the HV gas.
Consideration of the uncertainties on $N$(\ion{S}{2}) and $N$(\ion{S}{3}) and the upper limit on $N$(\ion{S}{4}) yields the fairly strict upper limit log[$N$(\ion{H}{1}) + $N$(\ion{H}{2})] (cm$^{-2}$) $\la$ 17.96.
The uncertainties in the total $N$(H) (and in the resulting depletions of C, N, and S) are dominated by the uncertainties in $N$(\ion{C}{3}) and $N$(\ion{N}{3}) (both derived from strong lines in lower resolution spectra) and in the adopted solar reference abundances for the three elements (0.05--0.07 dex).
Carbon may be very slightly depleted (by roughly 0.1 dex) in the HV gas, but within the uncertainties could be undepleted (and we note that the ratio of carbon to sulfur is essentially solar).

We also have fairly stringent upper limits on high-velocity absorption from \ion{N}{1} and (especially) \ion{O}{1}; the limit log[$N$(\ion{H}{1})] (cm$^{-2}$) $\la$ 14.84 in Table~\ref{tab:h2abund} assumes that oxygen is undepleted.
The HV gas toward $\zeta$ Ori is thus primarily ionized, with $N$(\ion{H}{1})/$N$(\ion{H}{2}) $\la$ 0.001.
The lack of high-velocity absorption from the higher ions \ion{S}{4}, \ion{Si}{4} and \ion{C}{4} suggests that only a narrow range of ionization states is significantly populated for any element.

The HV column densities for \ion{Al}{2} + \ion{Al}{3}, \ion{Si}{2} + \ion{Si}{3}, and \ion{Fe}{2} + \ion{Fe}{3} all correspond to somewhat smaller \ion{H}{2} column densities, suggesting some depletion of Al ($-$0.82 dex), Si ($-$0.39 dex), and Fe ($-$0.97 dex), relative to C, N, and S (see discussion in \S~\ref{sec-ddust}).
Since \ion{C}{3} and \ion{N}{3} are present (and dominant) in the HV gas, there could be some \ion{Al}{4} and \ion{Fe}{4} as well, however, so that the depletions of aluminum and iron could be less severe.
For temperatures consistent with the observed ion ratios (see below), collisional ionization equilibrium models (e.g., Sutherland \& Dopita 1993) suggest that \ion{Al}{4} may well be significant, but not \ion{Fe}{4}.
The shock models described below, however, suggest that \ion{Fe}{4} might also be significant.

Given the definition for the line width parameter $b$, we may obtain estimates for both the temperature ($T$) and internal turbulent velocity ($v_{\rm t}$) of the individual HV components by comparing the $b$-values for absorption lines due to elements of different atomic weight. 
For component 2H, where the line widths are fairly well-determined for \ion{C}{2}, \ion{N}{2},  \ion{Mg}{2}, \ion{Si}{2}, and \ion{Fe}{2}, the individual $b$-values listed in Table~\ref{tab:h2comp} yield firm upper limits $T_{\rm max}$ $\la$ 12,000 K in most cases, and the ensemble of $b$-values is most consistent with $T$ $\sim$ 9000$\pm$2000 K and $v_{\rm t}$ $\sim$ 0.5$\pm$0.5 km s$^{-1}$.
Since the $b$-values for the other HV components are not as well determined, we will assume that all the HV components have $T$ $\sim$ 9000 K.
Such a temperature is significantly smaller than would be inferred from the various ion ratios listed in Table~\ref{tab:hvgprop}, if the HV gas were assumed to be in CIE.
The HV gas is therefore likely not in CIE, but appears to have cooled from an initially higher temperature more rapidly than it could recombine, perhaps after having been shocked; photoionization may also be significant (see discussions below).
As the $b$-values for the HV \ion{Si}{3} components can be larger than those for the corresponding \ion{Si}{2} components, at least some fraction of the higher ions may be at somewhat higher $T$ and/or more broadly distributed.

In this ionized HV gas, the excited state \ion{C}{2}* is populated primarily via collisions with electrons, and the ratio $N$(\ion{C}{2}*)/$N$(\ion{C}{2}) can be used to estimate $n_{\rm e}$ (and thus $n_{\rm H}$) from the equation of excitation equilibrium, since the temperature is known.
In the limit of small $N$(\ion{C}{2}*)/$N$(\ion{C}{2}),
$n_e$ = 0.186~$T^{0.5}$~[$N$(\ion{C}{2}*)/$N$(\ion{C}{2})] (Trapero et al. 1996; Welty et al. 1999).
For $T$ $\sim$ 9000 K, the $n_{\rm e}$ $\sim$ $n_{\rm H}$ derived for HV components 2H and 4H are both $\sim$ 0.1--0.2 cm$^{-3}$. 
The limits for components 1H and 5H are consistent with such values, but the upper limit for component 3H is a factor 2--4 smaller (Table~\ref{tab:h2prop}).
The thermal pressures for components 2H and 4H, as measured by 2$n_{\rm e}T$, are thus about 2000--4000 cm$^{-3}$K.
Estimating individual HV component $N$(\ion{H}{2}) from $N$(\ion{S}{2}) + $N$(\ion{S}{3}), and adjusting for the adopted slightly higher total $N$(\ion{H}{2}), the thicknesses of components 2H and 4H are each about 0.5--0.6 pc; component 3H appears to be somewhat thicker.

\subsection{Intermediate-Velocity Gas ($-$60 km s$^{-1}$ $\la$ $v_{\odot}$ $\la$ $-$10 km s$^{-1}$)}
\label{sec-resiv}

For a number of the singly and doubly ionized species, the column densities in the IV gas are typically about a factor of 3 higher than in the HV gas (Table~\ref{tab:h2abund}).
From $N$(\ion{S}{2}) + $N$(\ion{S}{3}) [and the limits on $N$(\ion{S}{1}) and $N$(\ion{S}{4})], we estimate that log[$N$(\ion{H}{1}) + $N$(\ion{H}{2})] (cm$^{-2}$) $\sim$ 18.44 for the IV components.
The corresponding sums for carbon and nitrogen are not as well determined (the \ion{C}{3} $\lambda$977 line is very strong; the \ion{N}{3} $\lambda$989 line is strong and blended with a \ion{Si}{2} line), but suggest very similar IV $N$(H) (and negligible depletion for carbon and nitrogen).

In contrast to the HV gas, some \ion{O}{1} is detected in the strongest IV components near $-$16 km s$^{-1}$.
The implied column density log[$N$(\ion{H}{1})] (cm$^{-2}$) $\sim$ 15.31 then yields a neutral fraction $N$(\ion{H}{1})/$N$(\ion{H}{2}) $\sim$ 0.001 for the IV gas, so that the IV gas is also predominantly ionized.
The $N$(\ion{X}{3})/$N$(\ion{X}{2}) ratios are generally within a factor of 2 of the corresponding ratios in the HV gas (Table~\ref{tab:hvgprop}), suggesting that the degree of ionization is similar for both.

The IV column densities for \ion{Al}{2} + \ion{Al}{3} suggest that aluminum may be depleted by about $-$0.86 dex, relative to sulfur, in the IV gas --- close to the aluminum depletion inferred for the HV gas.
Unfortunately, the IV column densities for \ion{Si}{3} and \ion{Fe}{3} are not well determined --- the \ion{Si}{3} $\lambda$1206 line is saturated over part of the IV range and the \ion{Fe}{3} $\lambda$1122 line profile is noisy and incomplete --- so that we do not have strong constraints on the depletions of silicon and iron in the IV gas.
As was noted for the HV gas, the depletion of aluminum in the IV gas would be less severe if some \ion{Al}{4} is present.

The $b$-values adopted for most of the IV \ion{C}{2}, \ion{Mg}{2}, \ion{Al}{2}, \ion{Si}{2}, and \ion{Fe}{2} components suggest that the temperatures are comparable to those in the HV gas (i.e., about 9000 K).
Analysis of the \ion{C}{2} fine-structure excitation, assuming $T$ $\sim$ 9000 K, yields $n_e$ $\sim$ 0.16 cm$^{-3}$, and thus 2$n_e$$T$ $\sim$ 2500--3000 cm$^{-3}$K, for several of the stronger IV components --- very similar to the densities and thermal pressures found for the HV gas (Table~\ref{tab:h2prop}).
The IV components have total $N$(\ion{H}{2}) higher by a factor of about 3, however, and so the total thickness ($\sim$ 5.1 pc) is correspondingly larger.

While the lower ionization stages are generally dominant in the IV gas, some of the absorption from the higher ions \ion{S}{4}, \ion{Si}{4}, \ion{C}{4}, and (perhaps) \ion{N}{5} may be associated.
The ratio $N$(\ion{C}{4})/$N$(\ion{Si}{4}) = 7.0$\pm$0.9 found for the IV gas is significantly higher than the median value 3.8$\pm$1.9 for 31 lines of sight observed with {\it IUE} and/or GHRS (Sembach, Savage, \& Tripp 1997),
but is similar to the ratio found at intermediate velocities toward 23 Ori (Welty et al. 1999).
The GHRS echelle spectra suggest that the broad, weak absorption from \ion{C}{4} and \ion{Si}{4} toward $\zeta$ Ori arises from several narrower components, some at velocities similar to those found for components in lines of lower ionization stages such as \ion{Al}{3}, \ion{S}{3}, and \ion{Si}{3} (e.g., Savage, Sembach, \& Cardelli 1994; Fitzpatrick \& Spitzer 1997).

\section{Discussion}
\label{sec-disc}

\subsection{Orion's Cloak and Other High-Velocity Gas in the Galactic Disk}
\label{sec-dcloak}

Comparisons among the 21 cm, H$\alpha$, soft X-ray, and 100 $\mu$m emission maps of the Orion-Eridanus region have suggested the presence of a large, expanding shell, predominantly neutral but with an inner ionized zone, encompassing a bubble of hot gas at $T$ $\sim$ 10$^{6.2}$ K (Heiles 1976; Reynolds \& Ogden 1979; Burrows et al 1993; Brown et al. 1995; Snowden et al. 1995).
Cowie et al. (1979) ascribed the high-velocity gas seen with {\it Copernicus} toward a number of Orion stars (including $\zeta$ Ori) to a radiative shock due to one or more recent supernovae in that region.
They estimated a radius of about 120 pc (i.e., outside the expanding neutral shell) and an age of about 3 $\times$ 10$^5$ yr for this apparently large scale feature, as well as a pre-shock ambient density of about 3 $\times$ 10$^{-3}$ cm$^{-3}$.
Ionized gas at intermediate velocities is also seen toward a number of the Orion stars, suggesting that the IV gas also is a large-scale, regional feature --- not intimately associated with the individual stars. 
The HV and IV gas is not seen toward all the stars in the region, however, suggesting that its distribution is somewhat patchy (see also Shore 1982; Sonneborn et al. 1988).
Cowie et al. found the IV gas to be more highly ionized than could be attributed to collisional ionization at those velocities, and concluded that the IV gas is photoionized by (but at some distance from) the Orion association stars.
 
Higher resolution GHRS and IMAPS spectra covering additional species have enabled a more detailed understanding of the composition and ionization of this high- and intermediate-velocity gas, and have yielded estimates for the temperatures and electron densities in the {\it individual} HV and IV components (Trapero et al. 1996; Welty et al. 1999; this paper).
In Table~\ref{tab:hvgprop}, we compare total hydrogen column densities, ion ratios, depletions, and average physical properties for the HV and IV gas seen toward $\zeta$ Ori, 23 Ori, and $\tau$ CMa (the latter somewhat removed from the Orion's Cloak region; see the Appendix).
In all three sightlines, the high-velocity gas appears to consist of 4--5 components, whose similar widths (for a given ion) suggest that the HV component temperatures are typically in the range 7000--11000 K in all three cases. 
The total HV \ion{H}{2} column densities are comparable toward 23 Ori and $\zeta$ Ori, but $N$(\ion{H}{2}) is smaller by a factor of about 6--7 toward $\tau$ CMa.
The available $N$(\ion{X}{3})/$N$(\ion{X}{2}) ratios suggest that the HV gas toward $\zeta$ Ori (O9.5 Ib) is more highly ionized than the HV gas toward 23 Ori (B1 V), which in turn is more ionized than the HV gas toward $\tau$ CMa (O9 Ib) --- so that the degree of ionization is not correlated with the spectral type of the background star.
The depletions of silicon and aluminum in the HV gas (relative to typically undepleted sulfur) are slightly more severe toward $\zeta$ Ori than toward 23 Ori, and are least severe toward $\tau$ CMa.
Carbon, which is typically depleted by about 0.4 dex in both diffuse and moderately dense disk clouds at lower velocities (Sofia et al. 1997), appears to be essentially undepleted in the HV gas toward all three stars.
The density and thermal pressure are higher by a factor of about 3 in the HV gas toward 23 Ori, compared to the other two sightlines.

The total intermediate-velocity \ion{H}{2} column density is about 3 times higher toward $\zeta$ Ori than toward 23 Ori; no IV components were detected toward $\tau$ CMa [except, perhaps, in {\it Copernicus} spectra of \ion{C}{3} (Cohn \& York 1977)].
Comparisons of the $N$(\ion{X}{3})/$N$(\ion{X}{2}) ratios between the HV and IV gas, toward both $\zeta$ Ori and 23 Ori, suggest that the ratios may be slightly higher in the IV gas (though the ratio for carbon may be slightly lower in the IV gas toward $\zeta$ Ori).
The depletions in the IV gas are similar to those found for the respective HV components toward both $\zeta$ Ori and 23 Ori (with less severe depletions toward 23 Ori).
The IV components toward 23 Ori have significantly higher densities and thermal pressures than those toward $\zeta$ Ori.

We may also compare the high- and intermediate-velocity gas observed toward $\zeta$ Ori and 23 Ori with gas found at comparable velocites in other regions.
For example, absorption at $|v_{\rm LSR}|$ $\ga$ 60 km s$^{-1}$ (both positive and negative velocities) has been observed toward a number of stars located behind the Vela supernova remnant.
Jenkins, Silk, \& Wallerstein (1976) analyzed {\it Copernicus} spectra of four stars, and found C, N, Si, and Fe to be essentially undepleted in the high-velocity gas at $-$180 km s$^{-1}$ toward HD 74455 and at $-$90 km s$^{-1}$ toward HD 75821.
Limited data for various ratios $N$(\ion{X}{3})/$N$(\ion{X}{2}) suggest that the HV gas toward Vela is not as ionized as the HV gas toward Orion.
Jenkins, Wallerstein, \& Silk (1984) examined {\it IUE} spectra of 45 stars in the Vela region, and found the depletions of Al, Si, and Fe in the HV gas to be less severe than $-$1.2 dex, $-$0.5 dex, and $-$0.55 dex, respectively (relative to S) --- and thus less severe than the corresponding depletions in lower velocity clouds in the same region ($-$2.2 dex, $-$0.65 dex, and $-$1.2 dex).
Raymond, Wallerstein, \& Balick (1991) compared UV absorption lines toward HD 72088 with emission lines from adjacent nebulosity, attributed the HV gas to a shock with $v_s$ $\sim$ 150 km s$^{-1}$ and a swept up column density of about 10$^{18}$ cm$^{-2}$ (comparable to the values found for the HV gas toward Orion), and surmised that carbon might be undepleted.
Jenkins et al. (1998), using {\it HST}/STIS spectra, compared the column densities of various species normally dominant in \ion{H}{1} regions for the gas at $-$70 km s$^{-1}$ $\la$ $v_{\odot}$ $\la$ $-$45 km s$^{-1}$ and 112 km s$^{-1}$ $\la$ $v_{\odot}$ $\la$ 130 km s$^{-1}$ toward HD 72089.
The low $N$(\ion{O}{1})/$N$(\ion{S}{2}) and $N$(\ion{N}{1})/$N$(\ion{S}{2}) ratios suggest that the gas is ionized (though the ratios are not quite as low as those in the HV gas toward Orion).
The depletions of Al, Si, Fe, and Ni appear to be very mild ($-$0.6 dex, $-$0.3 dex, $-$0.4 to $-$0.8 dex, and $-$0.5 dex, relative to sulfur) ---  comparable to the values found in the HV gas toward Orion [though $N$(\ion{Fe}{2})/$N$(\ion{S}{2}) is lower toward Orion, where \ion{Fe}{3} appears to be the dominant ionization state of iron].
Determination of various ion ratios and physical properties characterizing the HV gas toward the Vela SNR would enable more detailed comparisons with the HV gas toward Orion; new data are required.

\subsection{Ionization Corrections / Models of Ionized Gas}
\label{sec-ioncor}

Determinations of interstellar abundances and depletions are often based on observations of a single ionization state --- usually presumed to be the dominant one for the gas in question.
In predominantly neutral (\ion{H}{1}) gas, the first ion (\ion{X}{2}) is dominant for most of the more abundant elements (those with first ionization potential less than about 13.6 eV), so that depletions may be fairly accurately estimated via ratios such as $N$(\ion{X}{2})/$N$(\ion{S}{2}) or $N$(\ion{X}{2})/$N$(\ion{Zn}{2}), since both sulfur and zinc are typically nearly undepleted (e.g., Welty et al. 1997).
Such simple estimates may not yield accurate relative abundances or depletions, however, for lower column density clouds (e.g., $N$(H) $\la$ 10$^{19}$ cm$^{-2}$), for significantly ionized gas, and for clouds where the degree of ionization is not known (e.g., Sembach et al. 2000; Howk \& Sembach 1999).
Unfortunately, data for potentially significant higher ions --- especially at a spectral resolution high enough to compare component structures with those found for the lower ions --- often are not available.
Of the elements we have considered in this paper, for example, only aluminum has detectable lines longward of Lyman $\alpha$ from its second ion (\ion{Al}{3}). 
While lines from \ion{Si}{3} and \ion{S}{3} just shortward of Lyman $\alpha$ can be observed with {\it HST}, the permitted resonance lines for \ion{C}{3}, \ion{N}{2}, \ion{N}{3}, and \ion{S}{4} all lie below 1100 \AA.
For studies of the Galactic ISM, IMAPS spectra are available for a small number of fairly bright stars; otherwise, we must depend on lower resolution spectra obtained with {\it Copernicus} or with {\it FUSE}.
For studies of QSO absorption-line systems, the shorter wavelength lines from those higher ions are entangled in the Lyman $\alpha$ forest.
Ionization corrections therefore are often estimated from theoretical models, by assuming a particular cloud structure (including coexistence of various ions), a radiation field (or lack thereof), and either collisional or photoionization equilibrium.

Because we have column densities for both singly and doubly ionized C, N, Al, Si, S, and Fe toward $\zeta$ Ori, we may (1) estimate empirical ionization corrections for the HV gas and (2) make more extensive comparisons with the predictions of various models of ionized gas.
For all six elements, the doubly ionized form is dominant in the HV gas; the ratio $N$(\ion{X}{3})/$N$(\ion{X}{2}) ranges from about 1.4 for nitrogen to about 11 for iron (Table~\ref{tab:hvgprop}).
In Table~\ref{tab:ratios}, we compare the observed HV gas ion ratios $N$(\ion{X}{2})/$N$(\ion{S}{2}) and $N$(\ion{X}{3})/$N$(\ion{S}{3}) with the ratios of the total column densities $N$(X)/$N$(S) --- all relative to the corresponding solar ratios listed in column 2.
Because the second ions are dominant in the HV gas, the [\ion{X}{3}/\ion{S}{3}] ratios provide the best estimates for the overall abundance ratios; all are within $\pm$0.1 dex of the corresponding [X/S].\footnotemark\
In this case, however, the [\ion{X}{2}/\ion{S}{2}] ratios do nearly as well --- within $\pm$0.2 dex --- except for iron.
\footnotetext{The square braces denote abundance ratios relative to the adopted solar ratios:  [X/Y] = log(X/Y) $-$ log(X/Y)$_{\odot}$.}

In Figure~\ref{fig:rat2h}, we compare the observed HV gas ratios $N$(\ion{X}{3})/$N$(\ion{X}{2}) and $N$(\ion{X}{4})/$N$(\ion{X}{2}) (given by the filled circles in each panel) with the values predicted by several models for ionized gas:

\begin{itemize}

\item{Sembach et al. (2000) presented ion fractions for models of relatively low density photoionized gas calculated using the CLOUDY code (e.g., Ferland et al. 1998).
In those models, Ti, Cr, Mn, and Fe can be doubly ionized more readily than most other elements, so that the ratios $N$(\ion{X}{2})/$N$(\ion{S}{2}) yield underestimates of the relative total abundances X/S for those four elements.
For most other elements, the ratios $N$(\ion{X}{2})/$N$(\ion{Fe}{2}) correspondingly overestimate the true X/Fe.
The magnitudes of the differences depend somewhat on the shape of the radiation field, but can easily exceed factors of 2.
The HV gas toward $\zeta$ Ori is significantly more ionized than the gas in the Sembach et al. models, however, as the observed ratios $N$(\ion{X}{3})/$N$(\ion{X}{2}) (for C, N, Al, Si, S, and Fe) are all more than an order of magnitude larger than the values predicted for their ``standard composite'' model (open circles in top panel).}

\item{Gruenwald \& Viegas (1992) tabulated ion column densities for lines of sight through model \ion{H}{2} regions characterized by several representative densities, metallicities, and radiation fields.
For the models with solar metallicity, $n_{\rm H}$ = 10 cm$^{-3}$ (the lowest density considered), and an O4 star radiation field (open squares in top panel), most elements have $N$(\ion{X}{3})/$N$(\ion{X}{2}) $\ga$ 1.5 near the center (upper point), decreasing to $\sim$ 1 by a fractional radius $r$ of 0.7 (lower point); sulfur is systematically more ionized, however.
While the ion ratios for carbon and nitrogen are (at some radii) similar to those seen in the HV gas toward $\zeta$ Ori, the ratios for silicon and iron are lower and the ratio for sulfur is higher.
For $r$ = 0.7, the predicted $N$(\ion{X}{4})/$N$(\ion{X}{2}) are consistent with the limits found for the HV gas --- but are much larger than those limits for smaller $r$.
We note, however, that $\theta^1$ Ori A and C appear to have the earliest spectral types (O7 and O6) of any of the stars in the Orion region; most of the brighter stars in the region have types between O9 and B1 (e.g., Warren \& Hesser 1977; Wall et al. 1996).
It is therefore unlikely that the radiation field toward $\zeta$ Ori could be characterized by as early a spectral type as O4.
For the softer B0 star field (open triangles in top panel; single points are for $r$ = 0.0), the first ions are generally dominant throughout (again except for sulfur); the ion ratios are thus lower for all $r$ and all elements (except sulfur) than those seen in the HV gas.
The ratios predicted for a slightly harder field (e.g., O9) and $r$ $\sim$ 0.0 might be reasonably consistent with most of the observed ratios --- except for $N$(\ion{Fe}{3})/$N$(\ion{Fe}{2}) (too low) and $N$(\ion{Si}{4})/$N$(\ion{Si}{2}) (too high).
Comparisons among the various models suggest that the $N$(\ion{X}{3})/$N$(\ion{X}{2}) ratios would be somewhat lower for densities below 10 cm$^{-3}$.}

\item{Sutherland \& Dopita (1993) have computed models for gas in collisional ionization equilibrium.
The middle panel of Figure~\ref{fig:rat2h} compares the ion ratios predicted for CIE at $T$ = 20,000 K (open squares), 40,000 K (open triangles), and 80,000 K (open circles) with those in the HV gas (filled circles).
The CIE models are consistent with the observed ionization of C, N, and O for $T$ $\sim$ 50,000--80,000 K, and with the observed ionization of Mg, Al, Si, S, and Fe for $T$ $\sim$ 25,000--40,000 K.
As noted above, however, the observed line widths imply a much lower temperature (near 9000 K), so the HV gas must not be in CIE.}

\item{Even models of cooling gas (e.g., Shapiro \& Moore 1976; Schmutzler \& Tscharnuter 1993; bottom panel, open squares and open triangles, respectively), however, predict lower $N$(\ion{X}{3})/$N$(\ion{X}{2}) ratios at $T$ = 10$^4$ K than those observed in the HV gas --- perhaps reflecting the neglect of ionization via radiation (stellar and/or shock-related) in the models.
On the other hand, the ratios $N$(\ion{C}{4})/$N$(\ion{C}{2}) and $N$(\ion{Si}{4})/$N$(\ion{Si}{2}) at $T$ = 10$^4$ K predicted by Shapiro \& Moore (1976) are much higher than the observed upper limits --- though comparison with the results of Schmutzler \& Tscharnuter suggests that those predicted ratios will decline rapidly below $T$ = 10$^4$ K.}

\end{itemize}

Comparisons between the observed HV gas ion ratios and the predictions from models of diffuse photoionized gas, denser \ion{H}{2} regions, collisionally ionized gas (in equilibrium), isochorically cooling gas, and shocked gas (see next section) thus indicate that a wide range of ion ratios may be produced in those different physical situations --- dependent both on current physical conditions and on the history of the gas.
If sufficient diagnostics are available, those differences can lead either to an accurate understanding of the observed gas or to the realization that other (as yet unexplored) processes must be significant.
Of the models noted above, photoionization by an O9 star radiation field might yield the best match with the observed ion ratios [based on a crude interpolation between the results for the O4 and B0 star models of Gruenwald \& Viegas (1992)] --- if the field due to the ensemble of stars in the region can mimic that computed for $r$ $\sim$ 0.0, and if the much lower observed density (0.1 cm$^{-3}$ versus 10 cm$^{-3}$) does not significantly affect the predicted ratios.
While some of the models predict some ion ratios consistent with those observed for the HV gas, none of the models appears able to reproduce all of the observed ion ratios.

\subsection{Shock Models for High- and Intermediate-Velocity Gas}
\label{sec-dshock}

The presence of ionized gas at high and intermediate velocities relative to the LSR; the proximity of that gas to the numerous young, massive stars in Orion; evidence from other data for an expanding shell produced by supernovae and/or stellar winds; the relatively mild depletions of Si, Al, and Fe; and indications that the gas is not in CIE all suggest that shocks may have played a major role in the evolution of that gas.
The high-resolution, high S/N UV spectra of $\zeta$ Ori have yielded information on the line profiles, ion ratios, and physical conditions in the HV and IV gas which can provide good tests for theoretical models of interstellar shocks.
In particular, comparisons of predicted and observed ratios of different ionization states of a single element [e.g., $N$(\ion{Si}{3})/$N$(\ion{Si}{2})] avoid possible complications due to differences in depletions (interesting in their own right) which may arise in comparisons of ions of different elements [e.g., $N$(\ion{C}{4})/$N$(\ion{Si}{4})].
We have considered two sets of shock models to try to understand the origins of the HV and IV gas toward $\zeta$ Ori:  
(1) for the HV gas, a shock moving toward us at a velocity of order $-100$ km s$^{-1}$ --- corresponding to the Cowie et al. (1979) model, and
(2) for the IV gas, a shock moving away from us, produced when a flow of high-velocity gas encounters lower velocity material --- as proposed by Jenkins \& Peimbert (1996) to explain the systematic broadening and small velocity shifts, with increasing rotational level $J$, of the H$_2$ lines near $v$ $\sim$ 0 km s$^{-1}$ seen in the IMAPS spectra of $\zeta$ Ori.
In that second model, the HV gas is the pre-shock flow.

\subsubsection{Models for the High-Velocity Gas}
\label{sec-dsmod1}

Ion column densities and absorption line profiles were computed for radiative shock wave models using an updated version of the computer code described by Raymond (1979) and Cox \& Raymond (1985).  
In brief, the model parameters are shock speed ($v_s$); pre-shock density ($n_0$), ionization state, and magnetic field ($B_0$); and elemental abundances.  
Because of numerical limitations in the code, the pre-shock densities are taken to be an order of magnitude or more larger than would produce $n_e$ $\sim$ $n_{\rm H}$ = 0.1-0.2 cm$^{-3}$ in the \ion{C}{2} region (as derived above).  
As long as we scale the magnetic field and the radiation field appropriately
(as $n_0^{1/2}$ and $n_0$, respectively), however, this will have no effect at all on the resulting column densities or line profiles.
The shock jump conditions (e.g. McKee \& Hollenbach 1980) are used to find the density, temperature and magnetic field strength just behind the shock.  
The code then follows the time-dependent ionization and radiative cooling of the shock-heated gas; photoionization by radiation emitted during the cooling is included.
In some of the calculations presented here, we also include ionizing radiation
($W_{\rm rad}$) from the Orion stars, approximated by a diluted Kurucz (1979) 25,000 K model spectrum.
All the models used the abundances of Baldwin et al. (1991).

Models for absorption lines require an additional model parameter that can often be ignored in emission line models.  
When the shocked gas cools to about 1000 K, it ceases to emit significant radiation at optical and UV wavelengths, and it does not matter how much cooler material accumulates.  
For absorption line models, however, the integration must be cut off at some point.  
Fortunately, the cutoff column density, log $N(H)$ (cm$^{-2}$) = 17.88, is tightly constrained for the $\zeta$ Ori high-velocity component.

For a strong shock, the flow speed is 3/4 $v_s$ just behind the shock, and approaches $v_s$ as the cooling gas is compressed.  
At each point in the flow, we compute the opacities in some of the strong UV lines, assuming a thermal velocity distribution at the local temperature, shifted to the local bulk velocity.  
These contributions to the optical depth are summed to produce the predicted absorption line profiles for the shock along a line normal to the shock front.  
Viewing the flow at an angle $\theta$ from the shock normal will increase the column densities by a factor of sec~$\theta$ and compress the velocity scale by a factor of cos~$\theta$.

In Figure~\ref{fig:shock1}, the observed HV profiles are compared with those predicted by several models, whose parameters are listed in Table~\ref{tab:mod1}. 
Some predicted ion ratios are given in column 5 of Table~\ref{tab:hvgprop} and are plotted in Figure~\ref{fig:rat1h}.
Model A (dotted lines) represents a ``simple'' 95 km s$^{-1}$ shock; models B (short-dashed lines) and C (long-dashed lines) add (in succession) a stronger radiation field and a stronger magnetic field.
Increasing the stellar radiation field (model B) generally decreases the column densities of the various singly ionized species and enhances the higher ions.
Increasing the magnetic field (model C) has somewhat the opposite effect --- the column densities of the singly ionized species are enhanced, while those of \ion{C}{4} and (to a lesser extent) \ion{Si}{4} are reduced.
A magnetic field also limits the compression of the shocked gas and reduces the velocity offset between high- and low-ionization species.

The observed HV gas line profiles appear to be in qualitative agreement with the basic shock picture.  
A 95 km s$^{-1}$ shock coming towards the Earth will produce narrow absorption features at that speed in the lowest ions (see, e.g., the observed \ion{C}{2}, \ion{Si}{2}, and \ion{Mg}{2} profiles), with an asymmetric extension to higher (i.e., less negative) velocities --- as well as \ion{C}{3}, \ion{N}{3}, \ion{Si}{3}, and \ion{S}{3} column densities similar to those observed.  
The much weaker lines from higher ionization stages (e.g., \ion{C}{4}, \ion{Si}{4}) are predicted to be less blue-shifted by up to $v_s$/4 $\sim$ 24 km s$^{-1}$.
Because the ionization lags behind the temperature and velocity evolution of the shock, the shifts for the higher ions in these models are less than half that maximum value. 

There are several major differences between the observed and predicted profiles and ion ratios, however:

\begin{itemize}

\item{First, the predicted blue shift is not quite large enough to match the observations unless the shock speed exceeds 100 km s$^{-1}$ --- but such fast shocks produce too much \ion{C}{4} and \ion{Si}{4} absorption.  
That problem is easily solved, as we have no reason to believe that the shock is moving through stationary gas, and random velocities of 10--20 km s$^{-1}$ are common in the ISM.  
We therefore choose the speed for the pre-shock medium ($v_0$) that shifts the strong absorption components to $v$ $\sim$ $-$95 km s$^{-1}$. 
Alternatively, adding a strong pre-shock magnetic field would absorb some of the shock energy --- thus permitting a higher speed shock without creating higher ionization.  
A strong field would also reduce the shock compression, however, so that the predicted speeds would not increase.}

\item{Second, while model A predicts $N$(\ion{X}{3})/$N$(\ion{X}{2}) ratios for C, N, and S that are lower than those observed (by factors of order 2), the corresponding predicted ratios for silicon and iron are too high by factors of 4--5 (Figure~\ref{fig:rat1h}).
The disagreements between observed and predicted ion ratios are typically somewhat worse for models B and C.
All three models predict too much \ion{Si}{4}; models A and B predict too much \ion{C}{4}.
It is perhaps interesting that the models predict too much highly ionized material for the refractory elements silicon and iron.
Such an effect might be expected if a significant fraction of the observed material is liberated from grains by the shock.
The refractory elements are only gradually (but increasingly) sputtered from the grains as the gas cools (e.g., Jones et al. 1994) --- so they are still severely depleted in the highest temperature gas.
Therefore, (1) only a small fraction of those atoms would reach the highest ionization states predicted by the models, and (2) the atoms will be returned to the gas from the dust predominantly as lower ions.
For elements that are initially severely depleted, the return of even 10\% of the atoms to the gas phase (see \S~\ref{sec-ddust} below) could have noticeable effects on both the ion ratios and the line profiles.}

\item{Third, the profile widths in the model spectra are systematically of order 50\% broader than those of the strongest observed absorption component 2H (note that the model profiles have not been convolved with the relevant instrumental response functions in the figure).
This discrepancy may suggest that the post-shock gas has not cooled enough in the models by the time the total column density has reached the cut-off value.
While the difference in widths also could be due to observing the shock at an angle of about 45 degrees, the large scale nature of the HV gas in Orion would seem to suggest that the line of sight to $\zeta$ Ori should be roughly normal to the shock.
Furthermore, the corresponding increase in shock speed would produce too much \ion{C}{4} and \ion{Si}{4}.}

\item{Fourth, the models produce a tail toward lower speeds in the profiles of the singly ionized species, but it is far narrower than the 20 km s$^{-1}$ tails in the observed \ion{C}{2} and \ion{N}{2} profiles.
In the HV gas toward 23 Ori (Welty et al. 1999b) and $\tau$ CMa (appendix), the corresponding redward absorption appears to be due to distinct additional components, as we have assumed in fitting the HV gas profiles toward $\zeta$ Ori.
(We note, however, that the models indicate that shocks can produce noticeably asymmetric profiles, contrary to what we have assumed in fitting the observed spectra.)}

\end{itemize}

A number of interesting physical effects might contribute to the extended lower speed tails.  
Inhomogeneities in the density or magnetic field on a length scale smaller than the cooling length (which is enormous, given the low density) are probably most likely.  
Such inhomogeneities could affect the shock speed ($n_0$) or the shock compression ($B_0$).  
Detailed computation of these effects is beyond the capability of the 1-D steady flow models presented here, but we estimate that density fluctuations of 60\% could account for the observed overall 30 km s$^{-1}$ spread in component speeds.  
The required magnetic field fluctuations depend on the values of $n_0$ and $B_0$, but for the $n_0$ in Model A, small regions with $B_0$ $\sim$ 1 $\mu$G could provide the lower speed gas.

The low densities derived from the $N$(\ion{C}{2}*)/$N$(\ion{C}{2}) ratios in the HV components --- $n_e$ $\sim$ $n_{\rm H}$ $\sim$ 0.1--0.2 cm$^{-3}$ --- suggest that (1) the pre-shock densities must have been very low (perhaps of order 0.001 cm$^{-3}$), and (2) the shock may be better characterized as a slow mode (rather than a fast mode) MHD shock (e.g., Draine \& McKee 1993).
The radiative cooling time just behind the shock is of order 1500/$n_0$ years --- which can be longer than the expansion time for adiabatic cooling for such low pre-shock densities.
Again, computation of detailed line profiles for such shocks is beyond the capabilities of the models considered here.

\subsubsection{Models for the Intermediate-Velocity Gas}
\label{sec-dsmod2}

We have also compared the absorption-line profiles of the intermediate-velocity gas with shock models.  
The presence of weak \ion{C}{4} and \ion{Si}{4}, along with an offset of about 10 km s$^{-1}$ between the low and high ionization species, suggests a shock speed of about 90--100 km s$^{-1}$ (similar to the values explored for the models of the HV gas).  
Faster shocks produce too much \ion{C}{4}, \ion{Si}{4}, and \ion{S}{4}, while slower shocks produce none.
The various ion ratios $N$(\ion{X}{3})/$N$(\ion{X}{2}) and the (relatively low) density found for the IV gas also are similar to those found for the HV gas.
 
The IV gas differs from the HV, however, in that its column density is larger by about a factor of 3, its velocity is only a quarter that of the HV gas, and the sense of the offset between high and low ionization states is inverted. 
The latter two aspects suggest that the direction of the shock is reversed compared to that of the HV gas.  
If gas moving at 100 km s$^{-1}$ strikes a stationary dense cloud, the gas just behind the shock is compressed by a factor $q$ and moves toward the cloud at a speed (100/$q$) km s$^{-1}$.  
As the gas cools and $q$ increases, its speed decreases.

In Figure~\ref{fig:shock2}, we compare the predicted profiles for the models with initial flow velocities of $-$95 to $-$100 km s$^{-1}$ (Table~\ref{tab:mod1}) with those observed for the IV gas toward $\zeta$ Ori; the predicted ion ratios are given in column 8 of Table~\ref{tab:hvgprop} and are shown in Figure~\ref{fig:rat1i}.
Model D (dotted lines) is the ``basic'' one here; models E (short-dashed lines) and F (long-dashed lines) add (in succession) a stronger magnetic field and a stronger radiation field.

As for the HV gas, there is qualitative agreement between the predicted and observed line profiles.
In particular, for these shocks moving away from us, the asymmetric tail is now toward more negative velocities, with an apparent progression in the profile shapes from high to low ions (e.g., for \ion{N}{3}, \ion{N}{2}, and \ion{N}{1} in Fig.~\ref{fig:spec1}).
The higher ions have more prominent absorption in the asymmetric tail, while the lower ions are more concentrated near $v$ $\sim$ 0 km s$^{-1}$ --- as would be expected for post-shock gas that starts hot at the most negative velocity, then cools and collapses toward $v$ $\sim$ 0 km s$^{-1}$.
Similarly, there also appears to be a systematic increase with $v$ in $n_e$$T^{-1/2}$, as indicated by the ratios $N$(\ion{C}{2}*)/$N$(\ion{C}{2}) and $N$(\ion{N}{2}*)/$N$(\ion{N}{2}) --- consistent with cooling, compressing gas. 

Again, however, there are also significant differences between the observed and predicted profiles:

\begin{itemize}

\item{First, the predicted absorption shortward of about $-$30 km s$^{-1}$ in the strong lines of \ion{C}{2}, \ion{C}{3}, and \ion{Si}{3} is generally weaker than the observed absorption.
A lower compression in the shock (e.g., by a factor 2.6 instead of the canonical factor 4) --- perhaps due to a magnetic field --- might help to enhance the absorption at more negative velocities (Jenkins \& Peimbert 1996).
As for the HV gas, the apparent discrete components seen in the IV absorption for several species --- which are not seen in the smooth predicted profiles --- could be due to inhomogeneities in the pre-shock density and/or magnetic field.}

\item{Second, the predicted ion ratios do not agree with the observed ratios (Fig.~\ref{fig:rat1i}).
For models D and E, the predicted column densities of most singly ionized species are somewhat higher than the observed values (especially for silicon and magnesium), while the corresponding $N$(\ion{X}{3}) are lower than the observed values (generally by factors of at least 4).
Moreover, these two models predict fairly strong IV absorption from \ion{N}{1}, which is not observed.
The low $N$(\ion{X}{3})/$N$(\ion{X}{2}) ratios and the strong \ion{N}{1} both are likely due to having no stellar radiation field in those two models.
For model F, which includes a strong stellar field, most of the ion ratios are higher than those observed for the IV gas.
A somewhat weaker radiation field than that adopted in model F --- perhaps similar to that in model A --- might yield better overall agreement with the observed ion ratios.}

\item{Third, while models D and E predict total IV column densities for \ion{C}{4} and \ion{Si}{4} close to the observed values, the observed absorption covers a wider velocity range.
Model F predicts somewhat higher IV column densities for those higher ions.}

\end{itemize}

The most serious problem with the reverse shock scenario, however, is the nature of the pre-shock gas, which should be visible near $-$100 km s$^{-1}$ --- unless (by a remarkable coincidence) the shock happens to have passed through most of the high-speed gas.
One could imagine that the HV components are just clouds of gas expanding away from the Orion stars (rather than a shock), and that the IV gas is the post-shock flow following a reverse shock.
However, a shock capable of slowing and heating the gas as much as is required would also compress the gas by perhaps an order of magnitude --- but the $N$(\ion{C}{2}*)/$N$(\ion{C}{2}) ratios suggest essentially equal densities in the HV and IV gas.
In view of this and the other problems listed above, the IV absorption may not result from a shock such as those described in this paper.
The more complex switch-on, switch-off shock scenario (with compression as low as a factor 2.6) proposed by Jenkins \& Peimbert (1996) is, unfortunately, beyond the capabilities of these models.

The abundances and physical properties characterizing the gas near $v$ $\sim$ 0 km s$^{-1}$ (the ``low-velocity obstruction'' in this reverse shock scenario) are also of interest.
Fits to the profiles of \ion{Zn}{2} and \ion{S}{2} suggest that the components near $v$ $\sim$ 0 km s$^{-1}$ contain about 10\% of the neutral gas toward $\zeta$ Ori.
Weak absorption from \ion{N}{2}, \ion{Si}{3}, and \ion{S}{3} indicate that a small amount of ionized gas is present as well [with $N$(\ion{X}{3})/$N$(\ion{X}{2}) $\la$ 0.01 for X = Al, Si, and S].
Column densities for \ion{Si}{2}, \ion{Al}{2}, and \ion{Fe}{2} (relative to those of \ion{S}{2} and \ion{Zn}{2}) suggest depletions slightly more severe than those typically found in warm, diffuse Galactic disk clouds (e.g., Savage \& Sembach 1996; Welty et al. 1999) --- and only slightly less severe than those in the strongest neutral components near $v$ $\sim$ 25 km s$^{-1}$ (Table~\ref{tab:depori}; component group A vs. component group DE).
Analysis of the \ion{C}{1} fine-structure excitation equilibrium (Jenkins \& Shaya 1979) indicates that the thermal pressure ($n_{\rm H}T$) in that gas is about 3 $\times$ 10$^4$ cm$^{-3}$K --- more than an order of magnitude higher than the value found for the strongest neutral components (Welty, Hobbs, \& Morton 2002).
While the increased thermal pressure might be suggestive of gas that has been shocked, the moderately severe depletions would seem to indicate that the accompanying dust grains have not been seriously disturbed.

\subsection{Dust Survival in High- and Intermediate-Velocity Gas}
\label{sec-ddust}

It is now well known that the depletions of various refractory elements are generally less severe in gas at higher LSR velocities.
The initial indications of such variations in depletion came from observations of increased $N$(\ion{Ca}{2})/$N$(\ion{Na}{1}) ratios at higher $|v_{\rm LSR}|$ (Routly \& Spitzer 1952; Sembach \& Danks 1994). 
Since both \ion{Ca}{2} and \ion{Na}{1} are trace species in those relatively low column density clouds, differences in ionization could in principle produce similar effects, however. 
Observations of UV absorption lines from the dominant ions (in \ion{H}{1} gas) of various severely depleted and mildly depleted elements subsequently provided strong evidence that the depletions do indeed vary with velocity (e.g., Shull, York, \& Hobbs 1977; Jenkins et al. 1976, 1984).
The relatively low spectral resolution and/or S/N achievable with {\it Copernicus} and {\it IUE}, and (in many cases) the incomplete coverage of potentially significant ions for some elements, however, generally did not allow very precise characterization of the depletions.

The increase in the gas phase abundances of the refractory elements at higher $|v_{\rm LSR}|$ is usually attributed to the (partial) destruction of dust grains in interstellar shocks, primarily in the warm, relatively low density ``intercloud'' medium.
A number of papers have explored the various processes affecting dust grains subjected to interstellar shocks --- including shattering and vaporization of grains due to grain-grain collisions and thermal and non-thermal sputtering of grain surfaces due to gas-grain collisions (e.g., Barlow \& Silk 1977; Cowie 1978; Draine \& Salpeter 1979; Seab \& Shull 1983; Jones et al. 1994; Jones, Tielens, \& Hollenbach 1996).
Jones et al. (1994, 1996), for example, find that grain destruction (i.e., the return of atoms or ions to the gas phase) peaks at a swept up hydrogen column density of 10$^{17}$--10$^{18}$ cm$^{-2}$, where the gas has cooled to about 10$^4$ K (comparable to the values found for the HV gas toward $\zeta$ Ori and 23 Ori).
For a 100 km s$^{-1}$ shock, the fraction of silicate dust destroyed (primarily by sputtering) ranges from $\la$ 10\% for grains with initial radius $\la$ 200 \AA\ to about 30\% for grains with initial radius $\ga$ 1000 \AA.
In general, the models yield time scales for grain destruction of order a few times 10$^8$ years --- much shorter than the time scales estimated for the injection of significant quantities of new grains into the ISM (e.g., in the winds of late-type stars).
The abundances of Mg, Si, and Fe found for different interstellar environments (e.g., ``cold, dense cloud'' vs. ``warm, diffuse cloud''), may suggest that the time scale for the return of silicon to the gas phase is significantly shorter, however (Tielens 1998). 
In order to maintain the depletions inferred for ``typical'' low-velocity interstellar clouds, the dust grains must therefore be re-assembled in the ISM --- presumably via accretion of atoms and/or ions onto the grain cores that survived the shock(s) --- during periods when the gas and dust cycle through the denser clouds.

Determination of column densities for several significant ionization states of C, N, S, Al, Si, and Fe in the HV and IV components toward $\zeta$ Ori affords an opportunity to gauge directly how much dust has survived in that ionized, perhaps shocked gas --- with fewer (uncertain) corrections for unobserved ions required.
As discussed above (\S~\ref{sec-reshv}), the HV column densities of \ion{C}{2} + \ion{C}{3}, \ion{N}{2} + \ion{N}{3}, and \ion{S}{2} + \ion{S}{3} are consistent with solar relative abundances of C, N, and S and an \ion{H}{2} column density of about 7.6 $\times$ 10$^{17}$ cm$^{-2}$ (though carbon may be slightly depleted). 
The IV column densities are about a factor of 3 higher (\S~\ref{sec-resiv}).
The corresponding column density sums for Al, Si, and Fe yield depletions of roughly $-$0.8 to $-$0.9 dex, $-$0.3 to $-$0.4 dex, and $-$0.95 dex, respectively, when compared to the inferred HV and IV $N$(\ion{H}{2}), though (as noted above) the depletions of aluminum and iron would be less severe (by perhaps 0.3--0.5 dex) if \ion{Al}{4} and \ion{Fe}{4} are significant.
The residual depletions in the HV and IV gas toward $\zeta$ Ori are less severe than those found for warm, diffuse Galactic clouds, but may be slightly more severe than those found for clouds in the Galactic halo (e.g., Savage \& Sembach 1996; Fitzpatrick 1996; Welty et al. 1999).

If the abundances in the HV components are representative of post-shock gas, then we must estimate the abundances in the pre-shock gas in order to determine how much of the dust was destroyed in the shock.
It is unlikely that the depletions in the pre-shock gas were as severe as those found in cold, dense clouds, given the relatively low $n_{\rm H}$ and $N$(H).
It would not be unreasonable, however, to adopt the depletions characterizing warm, low-density clouds for the pre-shock gas, since very similar values are found for the low $N$(H) components seen at low velocities toward both $\zeta$ Ori [e.g., near 0 km s$^{-1}$ --- ``component 1'' of Jenkins \& Peimbert (1996) or component group A in Table~\ref{tab:depori}] and 23 Ori [the ``weak, low-velocity'' gas described by Welty et al. (1999)]. 
If those depletions (listed in Table~\ref{tab:depori}) are representative of the pre-shock material toward $\zeta$ Ori, then the shock appears to have returned of order 10\% of the Al, Si, and Fe (and roughly 60\% of the carbon) from the dust to the gas phase.
While the $\sim$ 10\% destruction percentage for Al, Si, and Fe is comparable to that predicted by the theoretical models for a 100 km s$^{-1}$ shock (Jones et al. 1994, 1996), the similar values for silicon and iron stand in contrast to those derived from comparisons of cold cloud and warm cloud depletions that suggest that silicon is returned to the gas faster than iron (Tielens 1998; Jones 1999). 
The milder depletions of aluminum and silicon in the HV gas toward 23 Ori suggest that an even higher fraction of the dust was destroyed there.

In light of the residual depletions of Al, Si, and Fe in the HV gas, the high gas-phase abundance of carbon (and the corresponding inferred high dust destruction percentage) is intriguing.
In lower velocity clouds, the depletion of carbon is mild ($\sim$ $-$0.4 dex, relative to solar reference abundances), but roughly invariant --- with no discernible dependence on average sightline density $<n_{\rm H}>$ or molecular fraction $f$(H$_2$) (over the somewhat limited ranges sampled so far) (Sofia et al. 1997). 
The resulting amount of carbon in the dust is consistent with that needed to produce the various features in extinction and emission that have been attributed to carbonaceous material (e.g., Mathis 1996).
The depletions of most other elements (including Al, Si, and Fe) are typically more severe, and exhibit a clear dependence on $<n_{\rm H}>$, on $f$(H$_2$), and also (evidently) on $|v_{\rm LSR}|$.
Models for the destruction of dust in shocks predict that the time scale for destruction of graphite grains is roughly 1.5--2.0 times as long as the corresponding time scale for silicate grains (e.g., Jones et al. 1994, 1996).
We would thus expect that silicon would be returned to the gas phase more rapidly and more completely than carbon --- but that does not appear to be the case for the HV gas toward $\zeta$ Ori (though it may be for the HV gas toward 23 Ori).

\section{Conclusions / Summary}
\label{sec-sum}

We have used near-UV spectra obtained with the {\it Hubble Space Telescope} GHRS together with far-UV spectra from IMAPS and {\it Copernicus} to study the abundances and physical conditions in the ionized high- and intermediate-velocity gas seen along the line of sight to the star $\zeta$ Ori A.
The combination of high spectral resolution (FWHM $\sim$ 3.3--4.5 km s$^{-1}$), high S/N (90--550 for GHRS echelle; 30--80 for IMAPS), and broad spectral coverage has enabled a much more detailed and accurate view of the HV and IV gas than was possible using spectra from {\it Copernicus} alone.
With 2$\sigma$ detection limits better than 1 m\AA\ in many cases, we have obtained HV column densities for both singly and doubly ionized C, N, Al, Si, S, and Fe, and also for \ion{Mg}{2} and \ion{C}{2}*; we also have significantly restrictive limits for a number of neutral and triply ionized species. 
In the IV gas, we also detect absorption from \ion{O}{1}, \ion{C}{4}, and \ion{Si}{4}.
The column density sums $N$(\ion{X}{2}) + $N$(\ion{X}{3}) for C, N, and S yield estimates for the total hydrogen column density $N$(\ion{H}{1}) + $N$(\ion{H}{2}); by comparison, the corresponding sums for Al, Si, and Fe then provide estimates for the depletions.
Limits (or values) for $N$(\ion{O}{1}) and $N$(\ion{N}{1}) yield estimates for the amount of neutral gas $N$(\ion{H}{1}).
Fits to the high-resolution line profiles suggest (at least) 5 components for the HV gas and 8 components for the IV gas; comparisons of line widths, for species of different atomic weight, allow estimates for the temperature and turbulent velocity in each component.
Once $T$ is known, analysis of the \ion{C}{2} fine-structure excitation provides estimates for the local densities ($n_e$ $\sim$ $n_{\rm H}$) in several of the components.

The HV gas ($-$105 km s$^{-1}$ $\la$ $v$ $\la$ $-$65 km s$^{-1}$) is predominantly ionized, with \\
$N$(\ion{H}{1})/$N$(\ion{H}{2}) $<$ 0.001 and $N$(\ion{X}{3})/$N$(\ion{X}{2}) ranging from 1.4 (nitrogen) to 11 (iron).
The total hydrogen column density $N$(\ion{H}{2}) is about 7.6 $\times$ 10$^{17}$ cm$^{-2}$.
The depletions [$-$0.8 dex (Al), $-$0.4 dex (Si), and $-$0.95 dex (Fe), relative to C, N, and S] appear to be intermediate between those found in Galactic halo clouds and in warm, diffuse disk clouds --- though the depletions for aluminum and iron would be less severe if significant amounts of \ion{Al}{4} and \ion{Fe}{4} are present.
While the various $N$(\ion{X}{3})/$N$(\ion{X}{2}) ratios would be consistent with collisional ionization equilibrium at temperatures of 25,000--80,000 K, the component line widths imply $T$ $\sim$ 9000 K --- i.e., the HV gas is not in CIE.
Characteristic densities for the HV components are $n_e$ $\sim$ $n_{\rm H}$ $\sim$ 0.1--0.2 cm$^{-3}$; typical thermal pressures are 2$n_{\rm H}T$ $\sim$ 2000--4000 cm$^{-3}$K.

For the IV gas ($-$60 km s$^{-1}$ $\la$ $v$ $\la$ $-$10 km s$^{-1}$), the overall column densities are all roughly 3 times higher than for the HV gas, but the ion ratios, depletions, and inferred physical conditions are all similar to those found for the HV gas.

We have considered two sets of shock models to try to understand the origins of the HV and IV gas toward $\zeta$ Ori: 
(1) for the HV gas, a shock moving toward us at a velocity of order $-100$ km s$^{-1}$, and
(2) for the IV gas, a shock moving away from us, produced when a flow of high-velocity gas encounters lower velocity material.
The models are 1-D, assume steady flow, and include the effects of magnetic fields, stellar ionizing radiation, and radiation emitted as the gas behind the shock cools. 
The inclusion of a stellar radiation field seems necessary to reproduce the observed ion ratios $N$(\ion{X}{3})/$N$(\ion{X}{2}) and the weakness of \ion{N}{1}.
While the models can yield ion abundances and absorption-line profiles that resemble those observed, there are some significant differences.
For example, the models have difficulty reproducing the full set of observed ion ratios; the broad, complex tails in the observed profiles; and the general weakness of the observed \ion{C}{4} and \ion{Si}{4}.
Inhomogeneities in the density and/or magnetic field may produce the more complex structure seen in the observed profiles.
More complex shock models (e.g., for slow mode MHD shocks and/or switch-on, switch-off shocks) may be required to more accurately reproduce the observed line profiles and ion ratios.

We have also briefly examined other models for ionized gas --- diffuse photoionized gas, denser \ion{H}{2} regions, collisionally ionized gas in equilibrium, and isochorically cooling gas.
Those diverse models predict very different values for the various ion ratios [e.g., $N$(\ion{X}{3})/$N$(\ion{X}{2})]. 
While photoionization by an O9 star radiation field might yield a reasonable match with a number of the HV gas ion ratios, none of the various models appears able to reproduce the full set of ratios observed in the HV or IV gas toward $\zeta$ Ori.

If we assume that the depletions in the pre-shock gas were similar to those found in warm, diffuse disk clouds --- such as some of those seen at lower velocities toward $\zeta$ Ori and 23 Ori --- then the residual depletions of Al, Si, and Fe found in the HV and IV gas suggest that of order 10\% of the dust might have been destroyed in the shocks that produced the HV and IV gas.
While the $\sim$ 10\% destruction for Al, Si, and Fe is comparable to the percentages predicted by the theoretical models for a 100 km s$^{-1}$ shock, the similar values for silicon and iron disagree with expectations derived from comparisons of cold cloud and warm cloud depletions --- which suggest that silicon is returned to the gas faster than iron.
Moreover, the observed near-solar gas phase abundance of carbon appears inconsistent with the predicted longer time scales for the destruction of graphite grains.

\acknowledgments

Many of the observations reported here were obtained with the NASA/ESA {\it Hubble Space Telescope} at the Space Telescope Science Institute, which is operated by the Association of Universities for Research in Astronomy, Inc., under NASA contract NAS5-26555.
DEW is grateful for support provided by NASA grant GO-05896.01-94A (administered by STScI) and by NASA LTSA grant NAG5-3228.
Support for flying IMAPS on the ORFEUS-SPAS-I mission was provided by NASA grant NAG5-616 to Princeton University.
The ORFEUS-SPAS project was a joint undertaking of the US and German space agencies NASA and DARA.

\appendix

\section{High-Velocity Gas toward $\tau$ CMa}

During Cycle 5, we also obtained new GHRS ECH-A spectra of the O9 Ib star $\tau$ CMa ($V$ = 4.40; $E(B-V)$ = 0.13; $d$ $\sim$ 1 kpc), supplementing the ECH-B and G160M spectra described by Trapero et al. (1996), in order to probe in more detail the high-velocity gas along a line of sight somewhat removed from the Orion-Eridanus region, at $(l,b)$ = $(238\fdg2, -5\fdg5)$.
Six of the ECH-A settings listed in Table~\ref{tab:ghrs} (centered near 1190, 1206, 1260, 1334, 1402, and 1670 \AA) were used, and the resulting spectra were processed similarly to those obtained for $\zeta$ Ori.
Since $\tau$ CMa is significantly fainter than $\zeta$ Ori, the S/Ns in the final combined, normalized spectra were lower, ranging from about 55 to 125.

The normalized echelle spectra of species showing HV absorption, including the ECH-B spectrum of \ion{Mg}{2} $\lambda$2796 from Trapero et al. (1996), are shown in Figure~\ref{fig:tcma}.
In the figure (as in this paper) we focus on the high-velocity gas.
Table~\ref{tab:tcma} lists the equivalent widths or 2$\sigma$ upper limits for the high-velocity gas absorption for the transitions measured toward $\tau$ CMa (including some from Trapero et al.).
The last column in the Table gives the column density derived for each species, either from fits to the observed line profiles or, for upper limits, from the weak line approximation.
There does not appear to be significant absorption at intermediate velocities toward $\tau$ CMa, even in the strong \ion{C}{2} $\lambda$1334 and \ion{Si}{3} $\lambda$1206 lines.
We note, however, that IV absorption may be present for \ion{C}{3} toward $\tau$ CMa, and is present for \ion{C}{2}, \ion{C}{3}, and \ion{Si}{3} toward 29 CMa (24 arcmin away), in {\it Copernicus} spectra displayed by Cohn \& York (1977).

Table~\ref{tab:tcomp} lists the component parameters derived from fits to the stronger HV absorption features of \ion{C}{2}, \ion{Mg}{2}, \ion{Si}{2}, and \ion{Si}{3}.
The three first ions are likely to be very similarly distributed, and so their line profiles were fitted simultaneously.
The resulting velocity structure was used to fit the \ion{Si}{3} line at 1206 \AA, though there may be slight relative shifts for several of the \ion{Si}{3} components.

While we do not have as complete a set of column densities as for $\zeta$ Ori, the significantly smaller $N$(\ion{Si}{3})/$N$(\ion{Si}{2}) ratio for the HV gas toward $\tau$ CMa (especially in component 1) suggests that neither \ion{C}{3} nor \ion{N}{3} is likely to be significant.
We may thus use the observed log[$N$(\ion{C}{2})] = 13.55 cm$^{-2}$ to estimate log[$N$(\ion{H}{2})] $\sim$ 17.0 cm$^{-2}$ --- nearly an order of magnitude smaller than toward $\zeta$ Ori and 23 Ori.
The sum of the HV column densities of \ion{Si}{2} and \ion{Si}{3} imply that log[$N$(\ion{H}{2})] $\ga$ 16.96 cm$^{-2}$, while the corresponding sum for \ion{Al}{2} and \ion{Al}{3} yields 16.64 cm$^{-2}$ $\la$ log[$N$(\ion{H}{2})] $\la$ 17.05 cm$^{-2}$.
We thus find that silicon is essentially undepleted in the HV gas toward $\tau$ CMa (as found for the HV gas toward 23 Ori) and that aluminum is depleted by less than about 0.4 dex.
Furthermore, $N$(\ion{Mg}{2}) (alone) suggests that  log[$N$(\ion{H}{2})] $\ga$ 16.64 cm$^{-2}$, so that magnesium is also depleted by less than about 0.4 dex and that $N$(\ion{Mg}{3})/$N$(\ion{Mg}{2}) is much smaller in the HV gas toward $\tau$ CMa than toward the two Orion stars.

The $b$-values obtained in the profile fits to the \ion{C}{2}, \ion{Mg}{2}, and \ion{Si}{2} lines suggest that the temperatures of the individual components are similar to those in the HV gas toward $\zeta$ Ori and 23 Ori (i.e., 7000-9000 K).
Since we have only an upper limit for $N$(\ion{C}{2}*), we may only derive corresponding limits for the local densities, pressures, and thicknesses of the HV components --- which are generally consistent with the corresponding values found for the HV gas toward 23 Ori and $\zeta$ Ori (Table~\ref{tab:hvgprop}).

\newpage

\newpage

\begin{figure}
\figurenum{1b}
\epsscale{0.35}
\plotone{f1c.eps}
\caption{}
%\caption{Spectra of UV absorption lines observed toward $\zeta$ Ori.
%Tick marks denote the components derived from fits to the spectra.
%The vertical dotted lines separate the high-velocity (HV), intermediate-velocity (IV), and low-velocity (LV) component groups.
%Absorption due to other species is noted above several of the spectra; the dotted line in the S III spectrum shows the absorption due to Si II.
%Note the expanded vertical scale for S IV; where two lines are plotted for the HV gas (S II, S III, Fe II), the lower has been expanded by a factor of 5.}
\label{fig:spec2}
\end{figure}

\newpage

\begin{figure}
\figurenum{2}
\epsscale{0.4}
\plotone{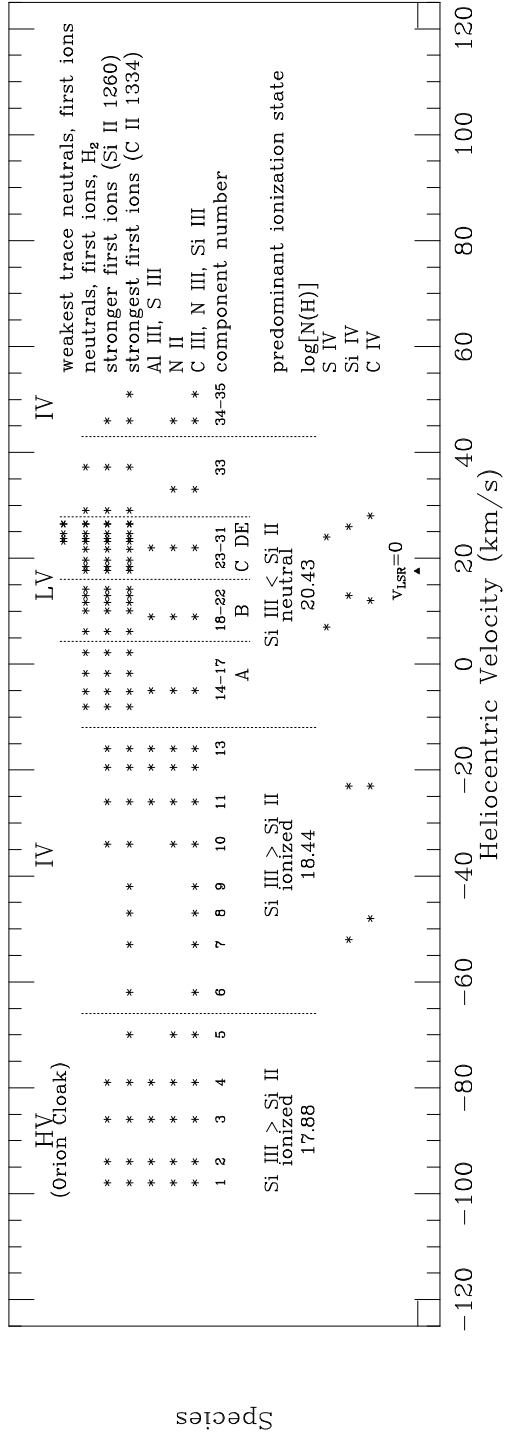}
\caption{Schematic diagram of the interstellar absorption observed toward $\zeta$ Ori.
The asterisks denote the individual components found in fits to the line profiles of the various species noted at the right.
The high- and intermediate-velocity gas (HV and IV) is predominantly ionized; the low-velocity gas (LV) is predominantly neutral.}
\label{fig:schem}
\end{figure}

\newpage

\begin{figure}
\figurenum{3}
\epsscale{0.35}
\plotone{f3.eps}
\caption{Schematic diagram showing column densities of ionization states of various elements observed in the high- and intermediate-velocity gas toward $\zeta$ Ori, arranged versus ionization potential.
Vertical dotted lines at 14.5 and 33.5 eV show the limits for species not detected in HV gas (N I has the highest $\chi_{\rm ion}$ for any undetected neutral species; Si III has the lowest $\chi_{\rm ion}$ for any species with undetected next higher ion); O I, C IV, and Si IV are detected in IV gas, however.}
\label{fig:abund}
\end{figure}

\newpage

\begin{figure}
\figurenum{4}
\epsscale{0.6}
\plotone{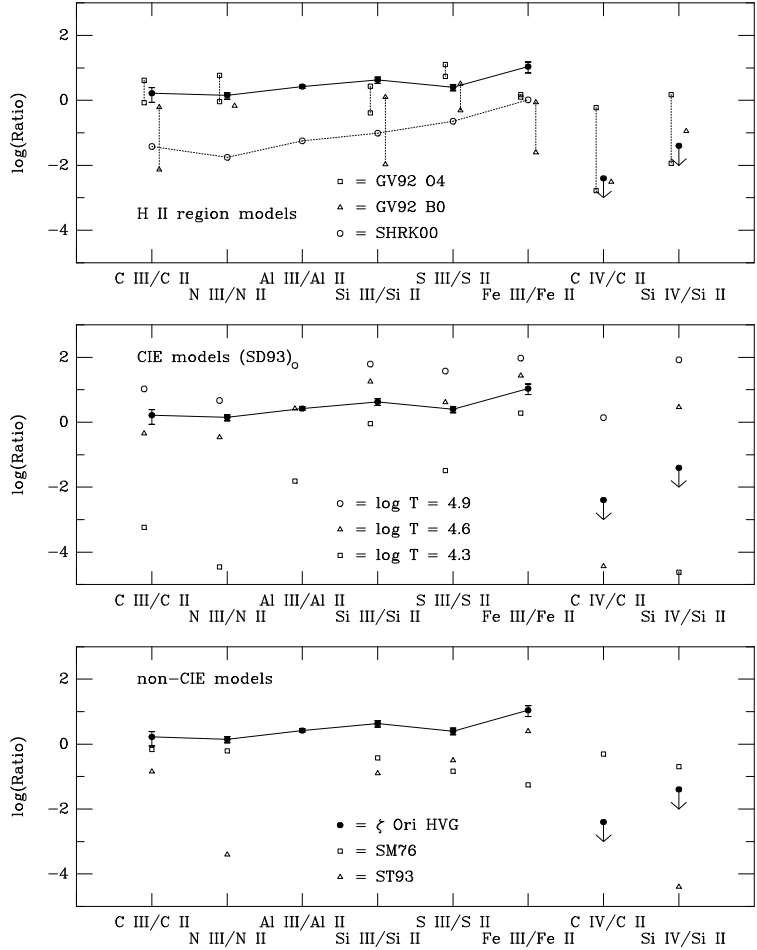}
\caption{Ion ratios observed for HV gas toward $\zeta$ Ori (filled circles; with 1-$\sigma$ uncertainties), compared with predictions from various models of ionized gas.
The top panel shows the ratios predicted by Gruenwald \& Viegas (1992) for H II regions ionized by an O4 star radiation field (open squares) or by a B0 star field (open triangles).
In each case, the upper point represents a sightline through the center of the region (fractional radius $r$ = 0.0); the lower is for a sightline at $r$ = 0.7; single points are for $r$ = 0.0.
The open circles show the values predicted for the ``standard composite'' model of Sembach et al. (2000) for diffuse ionized gas.
The middle panel shows the ratios predicted by the collisional ionization equilibrium models of Sutherland \& Dopita (1993), for $T$ = 20,000, 40,000, and 80,000 K.
The bottom panel shows the ratios predicted by two models of cooling (non-CIE) gas. 
The open squares give the ratios from Shapiro \& Moore (1976); the open triangles give values from Schmutzler \& Tscharnuter (1993) --- both for $T$ = 10$^4$ K (the lowest $T$ considered).}
\label{fig:rat2h}
\end{figure}

\newpage

\begin{figure}
\figurenum{5}
\epsscale{0.8}
\plotone{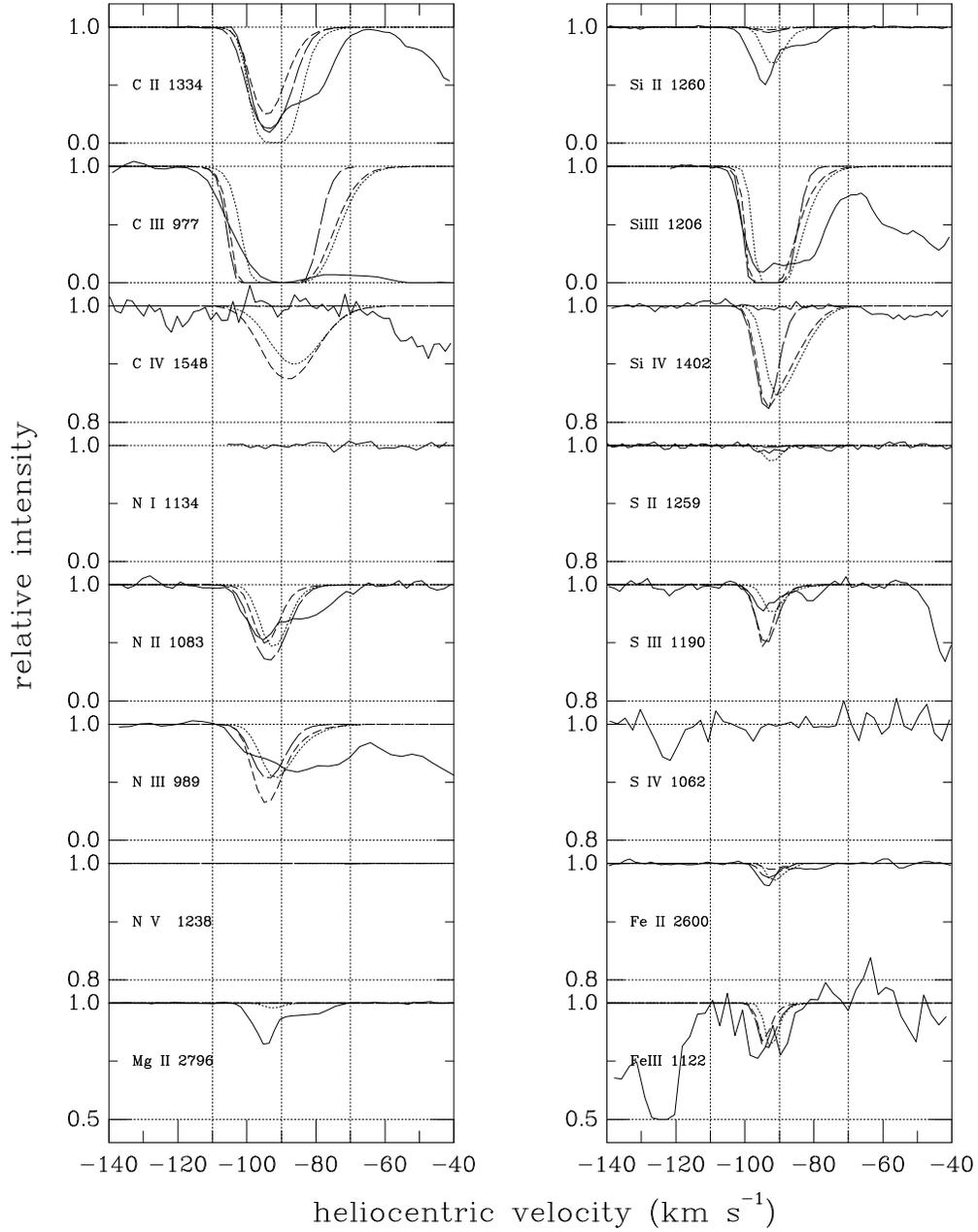}
\caption{Line profiles predicted by models of shocks moving toward us at 95--100 km s$^{-1}$, versus high-velocity line profiles observed toward $\zeta$ Ori (solid lines).
The models described in Table 8 (A, B, C) are given by dotted, short-dashed, and long-dashed lines, respectively.
Note the expanded vertical scale for some weak features.
N I and S IV are not computed in the models, but are expected to be weak.
The model profiles have not been convolved with the relevant instrumental response functions.}
\label{fig:shock1}
\end{figure}

\newpage

\begin{figure}
\figurenum{6}
\epsscale{0.8}
\plotone{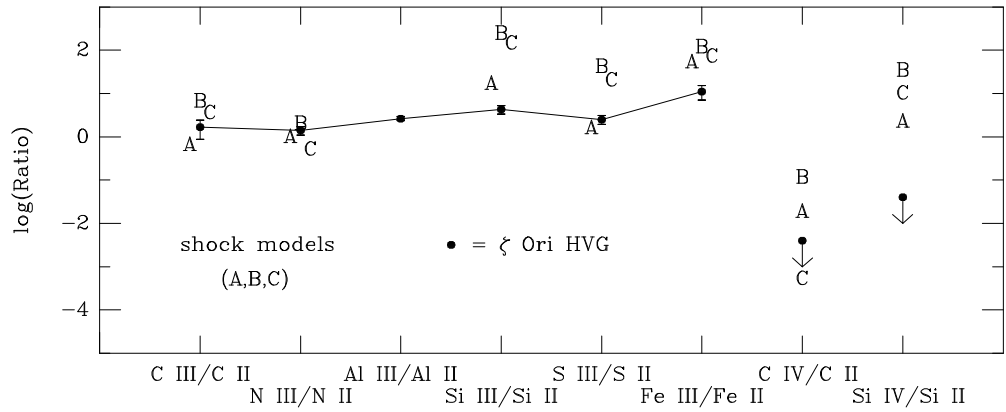}
\caption{Ion ratios observed for HV gas toward $\zeta$ Ori (filled circles; with 1-$\sigma$ uncertainties), compared with predictions from several shock models
(see Tables 6 and 8 and Figure 5).
Model A is the ``basic'' shock model; model B adds a stronger stellar radiation field; model C has both the stronger radiation field and a stronger magnetic field.}
\label{fig:rat1h}
\end{figure}
 
\newpage

\begin{figure}
\figurenum{7}
\epsscale{0.8}
\plotone{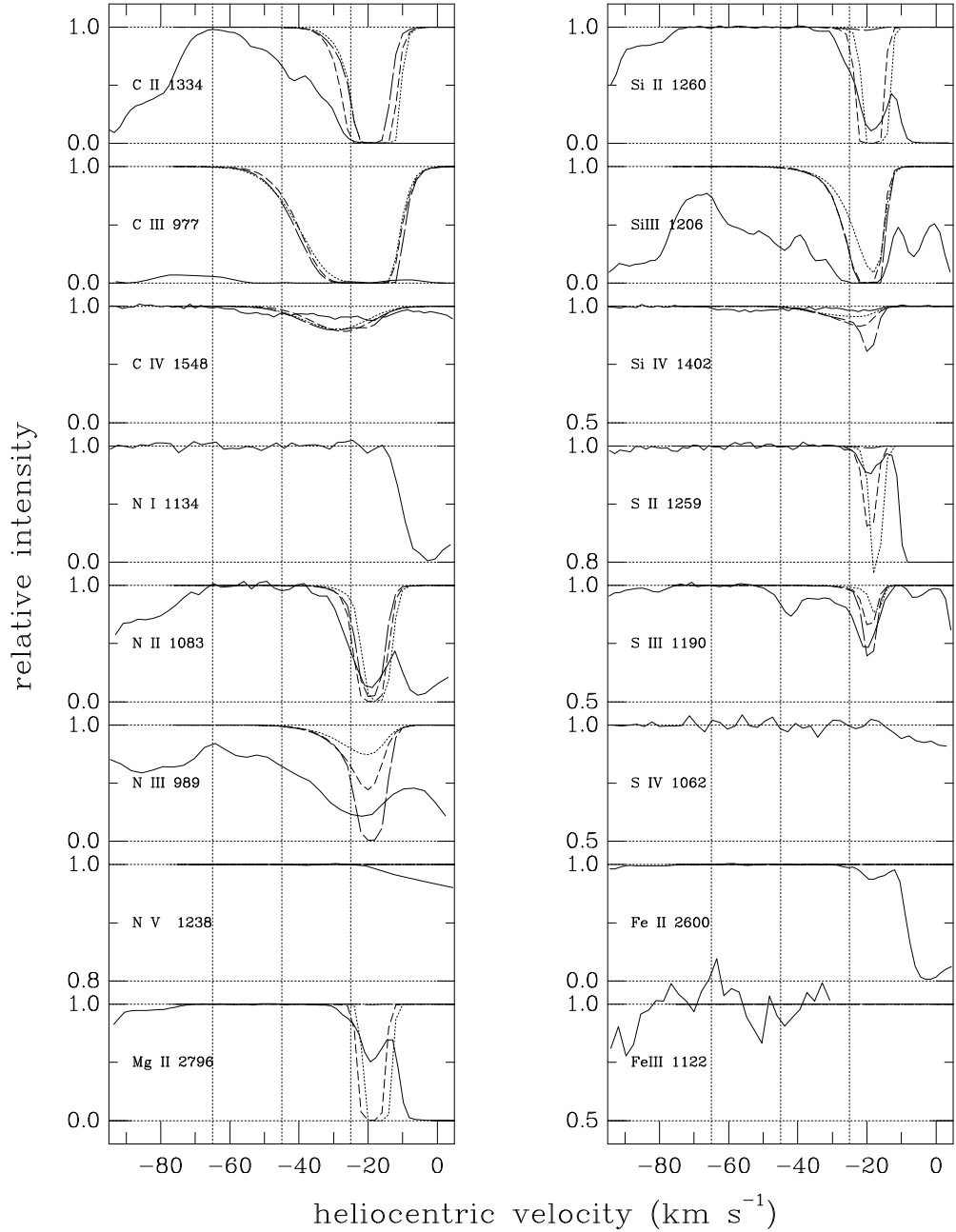}
\caption{Line profiles predicted by models of shocks moving away from us (produced by impact of high-velocity flow on low-velocity material), versus intermediate-velocity line profiles observed toward $\zeta$ Ori (solid lines).
The models described in Table 8 (D, E, F) are given by dotted, short-dashed, and long-dashed lines, respectively.
Note the expanded vertical scale for some weak features.
While profiles for N I and S IV are not computed in the models, the total
$N$(N I) is predicted to be much higher than observed for models D and E, but very low for model F; $N$(S IV) is very low in all three models.
Some of the absorption near IV S III is due to HV Si II.
The model profiles have not been convolved with the relevant instrumental response functions.}
\label{fig:shock2}
\end{figure}

\newpage

\begin{figure}
\figurenum{8}
\epsscale{0.8}
\plotone{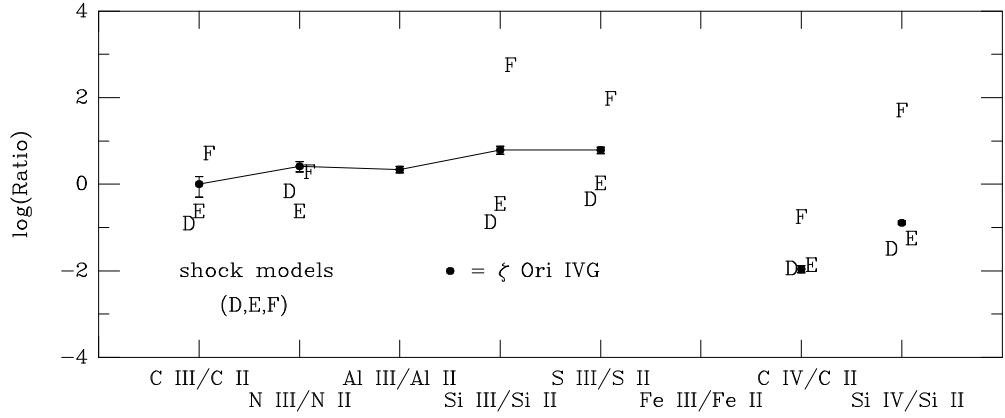}
\caption{Ion ratios observed for IV gas toward $\zeta$ Ori (filled circles; with 1-$\sigma$ uncertainties), compared with predictions from several (reverse) shock models
(see Tables 6 and 8 and Figure 7).
Model D is the ``basic'' shock model (with no stellar radiation field); model E adds a stronger magnetic field; model F has both the stronger magnetic field and a strong radiation field.}
\label{fig:rat1i}
\end{figure}

\clearpage

\begin{figure}
\figurenum{9}
\epsscale{0.45}
\plotone{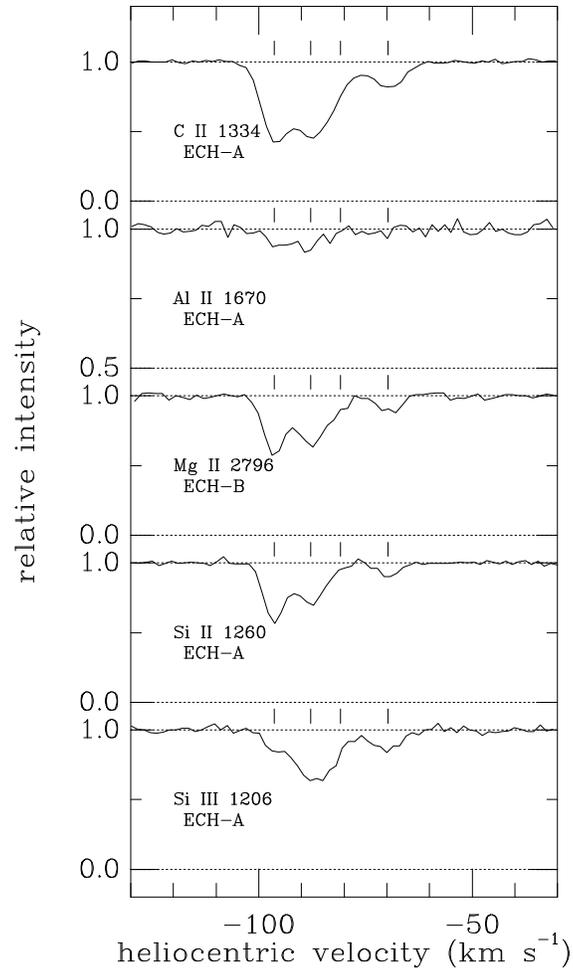}
\caption{High velocity absorption observed toward $\tau$ CMa.
Tick marks denote the components derived from fits to the spectra.
Note the expanded vertical scale for Al II.}
\label{fig:tcma}
\end{figure}

\clearpage

\begin{deluxetable}{ccccl}
\tabletypesize{\scriptsize}
\tablecolumns{5}
\tablenum{1}
\tablecaption{$\zeta$ Ori --- GHRS Spectra \label{tab:ghrs}}
\tablewidth{0pt}
\tablehead{
\multicolumn{1}{c}{Order}&
\multicolumn{1}{c}{$\lambda_{\rm min}$--$\lambda_{\rm max}$}&
\multicolumn{1}{c}{FWHM\tablenotemark{a}}&
\multicolumn{1}{c}{S/N}&
\multicolumn{1}{l}{Species\tablenotemark{b}}\\
\multicolumn{1}{c}{ }&
\multicolumn{1}{c}{(\AA)}&
\multicolumn{1}{c}{(km s$^{-1}$)}}
\startdata
\multicolumn{5}{c}{Echelle Spectra} \\
\tableline
A-47&1188.2--1194.8&3.56&130--210&Si II, Si II*, S III, C I, Cl I\\
A-47&1201.2--1207.5&3.40&115     &Si III\\
A-45&1257.5--1264.1&3.36&175--265&S II, Si II, Fe II, C I, C I*, C I**\\
A-43&1300.0--1307.2&3.53&235--285&P II, O I, S I, Si II\\
A-42&1330.4--1337.8&3.54&220--250&C II, C II*, Cl I, P III\\
A-40&1398.3--1406.1&3.53&270--280&Si IV, S I, Sn II\\
A-36&1545.3--1554.1&3.61& 90--245&C IV, Co II, Si I\\
A-34&1666.7--1675.5&3.35&105--120&Al II, P I, Mg I\\
B-31&1801.3--1811.5&3.56&195--270&Ni II, S I, Si II\\
B-30&1848.4--1859.1&3.66&190     &Al III, Fe I\\
B-27&2053.8--2065.7&3.67&315--415&Cr II, Zn II, Co II\\
B-25&2253.3--2265.5&3.43&455--525&Fe II, Al I, Ni I, Fe I\\
B-24&2334.4--2346.8&3.50&120     &Fe II (spybal)\\
B-22&2590.7--2604.0&3.27&385--555&Mn II, Fe II\\
B-20&2790.9--2806.7&3.55&415--500&Mg II, Mn I\\
\cutinhead{G160M Spectra\tablenotemark{c}}
\nodata&1186.6--1224.9&18.6&135--225&Si II, S III, C I, Mn II, N I, Si III, H I\\
\nodata&1205.7--1245.0&18.3&185--335&H I, Ge II, N V, Mg II\\
\nodata&1220.3--1258.7&18.0&180--345&Ge II, N V, Mg II, S II\\
\nodata&1239.5--1277.6&17.7&60--85  &Mg II, N V, S II, Si II, Fe II, C I, C I*, C I**, Si II*\\
\nodata&1315.4--1353.5&16.5&60--70  &C II, C II*, Cl I, C I, C I*, C I**, Ni II\\
\nodata&1367.7--1405.7&15.8&35--60  &Ni II, Si IV\\
\nodata&1382.3--1420.1&15.6&35--55  &Si IV, Ga II\\
\enddata
\tablenotetext{a}{For echelle spectra, resolution = 78000 + 6.4(m$\lambda$ $-$ 55000), where m is order number.  
For G160M spectra, resolution = 14400 + 16($\lambda$ $-$ 1100).}
\tablenotetext{b}{Possible; not all detected.}
\tablenotetext{c}{From HST archive; obtained by GO program 2584.}
\end{deluxetable}

\clearpage

\begin{deluxetable}{llcrrc}
\tabletypesize{\scriptsize}
\tablecolumns{6}
\tablenum{2}
\tablecaption{$\zeta$ Ori --- High-Velocity Gas Equivalent Widths \label{tab:ew}}
\tablewidth{0pt}

\tablehead{
\multicolumn{1}{c}{Ion}&
\multicolumn{1}{c}{$\lambda_{\rm vac}$\tablenotemark{a}}&
\multicolumn{1}{c}{ }&
\multicolumn{1}{c}{log(f$\lambda$)\tablenotemark{a}}&
\multicolumn{1}{c}{W$_{\lambda}$\tablenotemark{b}}&
\multicolumn{1}{c}{ }\\
\multicolumn{1}{c}{ }&
\multicolumn{1}{c}{(\AA)}&
\multicolumn{2}{c}{ }&
\multicolumn{1}{c}{(m\AA)}&
\multicolumn{1}{c}{ }}
\startdata
C I   &1193.0308 &E& 1.731 & $<$0.5       & \\
C II  &1036.3367 &I& 2.106 & 49.2$\pm$3.4 & \\
      &1334.5323 &E& 2.229 & 82.4$\pm$0.5 & \\
C II* &1335.6627 &E& 1.229 & \multicolumn{2}{c}{blend with C II*}\\
      &1335.7077 &E& 2.183 & 1.1$\pm$0.6  &B\\
C III & 977.026  &C& 2.872 & (161.5)      & \\
C IV  &1548.195  &E& 2.470 & $<$0.8       & \\
      &1550.770  &E& 2.169 & $<$0.4       & \\
&\\
N I   &1134.9803 &I& 1.691 & $<$2.0       & \\
      &1199.5496 &G& 2.193 & $<$1.5       & \\
N II  &1083.990  &I& 2.096 & 34.6$\pm$3.3 & \\
N III & 989.799  &C& 2.023 & (42.9)       & \\
N V   &1238.821  &G& 2.286 & $<$0.5       & \\
 &\\
O I   &1302.1685 &E& 1.831 & $<$0.4       & \\
 &\\
Mg I  &1668.4228 &E& 0.302 & $<$0.6       & \\
Mg II &2796.352  &E& 3.234 & 19.2$\pm$0.6 & \\
      &2803.531  &E& 2.933 &  9.2$\pm$0.7 & \\
 &\\
Al I  &2264.1637 &E& 2.305 & $<$0.2       & \\
Al II &1670.7874 &E& 3.486 & 4.3$\pm$1.4  & \\
Al III&1854.7164 &E& 3.017 & 4.3$\pm$1.0  & \\
 &\\
Si I  &1552.948  &E& 1.302 & $<$0.3       & \\
Si II &1260.4221 &E& 3.104 & 24.3$\pm$0.7 & \\
      &1304.3702 &E& 2.067 & 2.3$\pm$0.5  & \\
Si II*&1264.7377 &G& 3.058 & $<$2.4       & \\
Si III&1206.500  &E& 3.304 & [95.1]       & \\
Si IV &1402.770  &E& 2.554 & $<$0.25      & \\
 &\\
S I   &1807.3113 &E& 2.239 & $<$0.5       & \\
S II  &1259.519  &E& 1.311 & 0.7$\pm$0.4  & \\
S III &1190.208  &E& 1.421 & 2.1$\pm$0.7  & \\
S IV  &1062.671  &I& 1.628 & $<$1.7       & \\
 & \\
Ar I  &1066.6599 &I& 1.851 & $<$1.4       & \\
 & \\
Mn II &2594.499  &E&  2.847  & $<$0.5     & \\
 &\\
Fe I  &2260.2102 &E& 1.170 & $<$0.3       & \\
Fe II &2600.1729 &E& 2.765 & 2.9$\pm$0.4  & \\
Fe III&1122.526 &IC& 1.947 & 3$\pm$1      & \\
 &\\
\enddata
\tablenotetext{a}{Vacuum wavelengths and $f$-values from Morton (1991), updated by Welty et al. 1999.  E = GHRS echelle; G = GHRS G160M; I = IMAPS; C = {\it Copernicus}.}
\tablenotetext{b}{B = major contributor to blend.  Uncertainties and limits are 2-$\sigma$.  Values in parentheses are from Cowie et al. 1979.  Value for Si III in square braces is from profile fit.}
\end{deluxetable}

\clearpage

\begin{deluxetable}{cccccccccccc}
\tabletypesize{\tiny}
\tablecolumns{12}
\tablenum{3}
\tablecaption{$\zeta$ Ori --- HV, IV Component Structure\tablenotemark{a} \label{tab:h2comp}}
\tablewidth{0pt}
\tablehead{
\multicolumn{1}{c}{$v_{\odot}$}&
\multicolumn{1}{c}{C II}&
\multicolumn{1}{c}{C II*}&
\multicolumn{1}{c}{N II}&
\multicolumn{1}{c}{Mg II}&
\multicolumn{1}{c}{Al II}&
\multicolumn{1}{c}{Al III}&
\multicolumn{1}{c}{Si II}&
\multicolumn{1}{c}{Si III}&
\multicolumn{1}{c}{S II}&
\multicolumn{1}{c}{S III}&
\multicolumn{1}{c}{Fe II}\\
\multicolumn{1}{c}{(km s$^{-1}$)}&
\multicolumn{1}{c}{(10$^{13}$)}&
\multicolumn{1}{c}{(10$^{11}$)}&
\multicolumn{1}{c}{(10$^{12}$)}&
\multicolumn{1}{c}{(10$^{11}$)}&
\multicolumn{1}{c}{(10$^{10}$)}&
\multicolumn{1}{c}{(10$^{10}$)}&
\multicolumn{1}{c}{(10$^{11}$)}&
\multicolumn{1}{c}{(10$^{12}$)}&
\multicolumn{1}{c}{(10$^{11}$)}&
\multicolumn{1}{c}{(10$^{12}$)}&
\multicolumn{1}{c}{(10$^{10}$)}}
\startdata
-97.0$\pm$0.2 & 0.15$\pm$0.04 & $<$0.6 & 5.6$\pm$0.8 & 0.5$\pm$0.1 &
    0.6$\pm$0.5 & 2.2$\pm$1.1 & 2.4$\pm$0.2 & 0.28$\pm$0.16 & $<$3.1 & $<$0.3 & $<$1.9 \\
 1H& (3.5) & (3.5) & (3.3) & (2.2) & (2.1) & (2.1) & (2.0) & (2.5) & (1.9) & (1.9) & (1.5) \\
 & \\
$-$92.5$\pm$0.1 &  3.7$\pm$0.1 &  4.4$\pm$0.5 & 11.4$\pm$0.6 &  2.4$\pm$0.1 & 
    4.2$\pm$0.4 & 10.1$\pm$0.9 & 10.8$\pm$0.2 &  3.1$\pm$0.2 & 12.3$\pm$2.8 & 
    3.4$\pm$0.3 & 13.2$\pm$1.9 \\
2H & 3.8$\pm$0.4 & (3.8) & 3.4$\pm$0.5 & 2.5$\pm$0.4 & (2.5) & (2.5) & 
    2.4$\pm$0.3 & 3.7$\pm$0.2 & (2.3) & (2.3) & 2.1$\pm$0.4 \\
 & \\
$-$85.5$\pm$0.2 & 2.0$\pm$0.1 & $<$0.5      & 8.6$\pm$0.3 & 1.1$\pm$0.1 & 
    3.0$\pm$0.5 & 9.0$\pm$1.1 & 5.4$\pm$0.2 & 3.0$\pm$0.1 & 7.7$\pm$3.7 & 
    2.3$\pm$0.4 & 4.9$\pm$2.5 \\
 3H& (5.5) & (5.5) & (5.4) & (5.0) & (5.0) & (5.0) & (5.0) & (4.7) & (4.9) & (4.9) & (4.5) \\
 & \\
$-$77.5$\pm$0.2 & 1.6$\pm$0.1 & 0.9$\pm$0.4 & 8.7$\pm$0.4 & 0.7$\pm$0.1 & 
    1.7$\pm$0.4 & 3.3$\pm$0.9 & 2.4$\pm$0.1 & 2.2$\pm$0.2 & 6.3$\pm$3.0 & 
    2.2$\pm$0.3 & 3.7$\pm$2.0 \\
4H & (5.5) & (5.5) & (5.5) & (3.9) & (3.7) & (3.7) & (3.4) & 4.4$\pm$0.3 & (3.3) & (3.3) & (2.9) \\
 & \\
$-$67.7$\pm$0.5 & 0.06$\pm$0.01 & $\la$0.4 & 1.2$\pm$0.2 & $<$0.06 &
    $<$0.3 & $\la$0.8 & $<$0.08 & 0.5$\pm$0.1 & $<$2.4 & $<$0.3 & $<$1.6 \\
5H & (3.5) & (3.5) & (3.3) & (2.2) & (2.1) & (2.1) & (2.0) & (5.0) & (1.9) & (1.9) & (1.5) \\
 & \\
$-$58.8$\pm$0.9 & 0.06$\pm$0.01 & \nodata & \nodata & \nodata &\nodata & \nodata & \nodata &
     1.0$\pm$0.1 & \nodata & \nodata & \nodata \\
6I & (3.5) & \nodata & \nodata & \nodata & \nodata & \nodata & \nodata & 4.2$\pm$0.5 & \nodata & 
     \nodata & \nodata \\
 & \\
$-$51.8$\pm$0.5 & 0.20$\pm$0.05 & \nodata & \nodata & \nodata &\nodata & \nodata & \nodata &
     1.0$\pm$0.1 & \nodata & \nodata & \nodata \\
7I & 2.5$\pm$0.7 & \nodata & \nodata & \nodata & \nodata & \nodata & \nodata & (3.5) & \nodata & 
     \nodata & \nodata \\
 & \\
$-$45.9$\pm$1.0 & 0.36$\pm$0.04 & \nodata & \nodata & \nodata &\nodata & \nodata & \nodata &
     0.8$\pm$0.2 & \nodata & \nodata & \nodata \\
8I & (3.5) & \nodata & \nodata & \nodata & \nodata & \nodata & \nodata & (3.0) & \nodata & 
     \nodata & \nodata \\
 & \\
$-$39.9$\pm$0.2 & 0.8$\pm$0.1 & \nodata & \nodata & \nodata &\nodata & \nodata & \nodata &
     1.4$\pm$0.4 & \nodata & \nodata & \nodata \\
9I & 3.2$\pm$0.4 & \nodata & \nodata & \nodata & \nodata & \nodata & \nodata & 3.1$\pm$0.7 & \nodata & 
     \nodata & \nodata \\
 & \\
$-$32.0$\pm$0.4 & 1.6$\pm$0.1 & $<$5.0 & 1.3$\pm$0.4 & $<$0.04 &
     $<$0.7 & $<$1.3 & 1.3$\pm$1.0 & 2.2$\pm$0.1 & $<$4.8 &
     1.5$\pm$0.8 & $<$2.0 \\
10I & (4.2) & (4.2) & (4.0) & (2.0) & (2.0) & (2.0) & (2.0) & 3.5$\pm$0.4 & (2.0) & (2.0) & (1.9) \\
 & \\
$-$24.3$\pm$0.4  & 4.0$\pm$0.7 &[3.9$\pm$0.3]& (5.9$\pm$0.2) & 1.0$\pm$0.2 &
     2.7$\pm$1.0 & 2.8$\pm$0.8 & 5.0$\pm$1.0 & (3.1$\pm$0.2) & (2.5) &
     4.5$\pm$0.9 & 6.1$\pm$2.6 \\
11I & (3.3) & (3.3) & (3.0) & (2.5) & (2.5) & (2.5) & (2.5) & (2.5) & (2.5) & (2.5) & (2.3) \\
 & \\
$-$17.8$\pm$0.4 & 25.5$\pm$4.4 &[24.3$\pm$1.7]& (40.$\pm$2.) & 7.7$\pm$0.3 &
     19.5$\pm$3.8 & 61.$\pm$3. & 32.$\pm$2. & (18.7$\pm$1.0) & 48.$\pm$7. &
     28.$\pm$2. & 44.$\pm$7. \\
12I & (3.7) & (3.7) & (3.5) & (3.0) & (3.0) & (4.0) & (3.0) & (3.0) & (3.0) & 
     (3.0) & (2.8) \\
 & \\
$-$14.1$\pm$0.4 & 18.4$\pm$3.2 & [17.5$\pm$1.2] & (30.$\pm$1.) & 4.2$\pm$0.3 &
     13.5$\pm$4.4 & 16.$\pm$3. & 23.$\pm$3. & (9.4$\pm$0.5) & 15.4$\pm$7.5 &
     6.0$\pm$2.5 & 27.$\pm$8. \\
13I & (4.0) & (4.0) & (3.8) & (3.5) & (3.5) & (3.5) & (3.5) & (3.5) & (3.5) & (3.5) & (3.2) \\
 & \\
\enddata
\tablenotetext{a}{Entries are N ($\times$ the factor under the ion name, in cm$^{-2}$) and b (km s$^{-1}$).   HV, IV values in parentheses were fixed in the fitting; values in square brackets were varied together.  Uncertainties are 1-$\sigma$; limits are 2-$\sigma$.  Components (1--5) are HV; components (6--13) are IV.}
\end{deluxetable}

\clearpage

\begin{deluxetable}{lcc}
\tabletypesize{\scriptsize}
\tablecolumns{3}
\tablenum{4}
\tablecaption{$\zeta$ Ori --- HV, IV Column Densities and Depletions\tablenotemark{a} \label{tab:h2abund}}
\tablewidth{0pt}
\tablehead{
\multicolumn{1}{c}{Ion}&
\multicolumn{1}{c}{log[N(HV)]}&
\multicolumn{1}{c}{log[N(IV)]}\\
\multicolumn{1}{c}{ }&
\multicolumn{1}{c}{(cm$^{-2}$)}&
\multicolumn{1}{c}{(cm$^{-2}$)}}
\startdata
H I    &  $<$14.84               & 15.31$^{+0.08}_{-0.10}$  \\
H II   & 17.88$^{+0.05}_{-0.06}$ & 18.44$^{+0.05}_{-0.05}$  \\
 &\\
C I    & $<$11.9                 & $<$11.9 \\
C II   & 13.88$^{+0.01}_{-0.01}$ & 14.71$^{+0.05}_{-0.06}$  \\
C II*  & 11.79$^{+0.07}_{-0.09}$ & 12.66$^{+0.02}_{-0.02}$  \\
C III  & 14.1$^{+0.2}_{-0.2}$    & 14.7$^{+0.2}_{-0.2}$  \\
C IV   &  $<$11.47               & 12.75$^{+0.04}_{-0.04}$  \\
D(C) & $-$0.12$^{+0.13}_{-0.13}$ & +0.02$^{+0.10}_{-0.15}$  \\
 &\\
N I    &  $<$12.0                & $<$12.0  \\
N II   & 13.55$^{+0.01}_{-0.01}$ & 13.89$^{+0.01}_{-0.01}$   \\
N III  & 13.7$^{+0.1}_{-0.1}$    & 14.3$^{+0.1}_{-0.1}$  \\
N V    &  $<$11.4                & $<$11.4  \\
D(N)   & +0.08$^{+0.05}_{-0.05}$ & +0.03$^{+0.08}_{-0.10}$  \\
 &\\
O I    &  $<$11.71               & 12.18$^{+0.08}_{-0.10}$  \\
 &\\
Mg I   & $<$13.3                 & $<$13.3  \\
Mg II  & 11.67$^{+0.02}_{-0.02}$ & 12.11$^{+0.02}_{-0.02}$  \\
 &\\
Al I   &  $<$10.7                & $<$10.7  \\
Al II  & 10.98$^{+0.04}_{-0.05}$ & 11.55$^{+0.07}_{-0.07}$  \\
Al III & 11.40$^{+0.04}_{-0.03}$ & 11.90$^{+0.02}_{-0.02}$  \\
D(Al) &$-$0.82$^{+0.02}_{-0.03}$ & $-$0.86$^{+0.05}_{-0.05}$  \\
 &\\
Si I   &  $<$12.0                & $<$12.0  \\
Si II  & 12.32$^{+0.01}_{-0.01}$ & 12.79$^{+0.02}_{-0.03}$  \\
Si II* &  $<$11.3                & $<$11.3  \\
Si III & 12.95$^{+0.09}_{-0.10}$ & 13.57$^{+0.08}_{-0.09}$  \\
Si IV  &  $<$10.93               & 11.89$^{+0.02}_{-0.03}$  \\
D(Si)& $-$0.39$^{+0.08}_{-0.08}$ & $-$0.34$^{+0.08}_{-0.10}$ \\
 &\\
S I    &  $<$11.3                & $<$11.3  \\
S II   & 12.47$^{+0.09}_{-0.11}$ & 12.82$^{+0.06}_{-0.08}$  \\
S III  & 12.91$^{+0.03}_{-0.04}$ & 13.60$^{+0.03}_{-0.03}$  \\
S IV   & $<$12.65                & $<$12.65  \\
D(S)   &$-$0.11$^{+0.04}_{-0.03}$&$-$0.04$^{+0.05}_{-0.07}$  \\
 &\\
Fe I   &  $<$12.0                & $<$12.0  \\
Fe II  & 11.34$^{+0.08}_{-0.10}$ & 11.88$^{+0.05}_{-0.06}$  \\
Fe III & 12.38$^{+0.12}_{-0.17}$ & \nodata  \\
D(Fe)& $-$0.97$^{+0.11}_{-0.16}$ & \nodata  \\
\enddata
\tablenotetext{a}{Uncertainites are 1-$\sigma$; limits are 2-$\sigma$.  H I estimated from O I; H II estimated from C II + C III, N II + N III, S II + S III --- using reference abundances from Anders \& Grevesse 1989 and Grevesse \& Noels 1993.  Depletions are relative to those reference abundances.}
\end{deluxetable}

\clearpage

\begin{deluxetable}{cccccc}
\tabletypesize{\scriptsize}
\tablecolumns{6}
\tablenum{5}
\tablecaption{$\zeta$ Ori --- HV, IV Component Properties \label{tab:h2prop}}
\tablewidth{0pt}
\tablehead{
\multicolumn{1}{c}{Comp\tablenotemark{a}}&
\multicolumn{1}{c}{$N$(C II*)/$N$(C II)\tablenotemark{b}}&
\multicolumn{1}{c}{$n_e$ = $n_{\rm H}$\tablenotemark{c}}&
\multicolumn{1}{c}{log[$N$(H II)]\tablenotemark{d}}&
\multicolumn{1}{c}{r}&
\multicolumn{1}{c}{log(2$n_{\rm H}T$)}\\
\multicolumn{1}{c}{ }&
\multicolumn{1}{c}{ }&
\multicolumn{1}{c}{(cm$^{-3}$)}&
\multicolumn{1}{c}{(cm$^{-2}$)}&
\multicolumn{1}{c}{(pc)}&
\multicolumn{1}{c}{(cm$^{-3}$K)}\\}
\startdata
 1H & $<$0.04         & $<$0.7          & 16.5  & $>$0.015 & $<$4.3 \\
 & \\
 2H & 0.012$\pm$0.001 & 0.21$\pm$0.03   & 17.48 & 0.47     & 3.6 \\
 & \\
 3H & $<$0.003        & $<$0.05         & 17.36 & $>$1.5   & $<$3.0 \\
 & \\
 4H & 0.006$\pm$0.003 & 0.10$\pm$0.05   & 17.25 & 0.58     & 3.3 \\
 & \\
 5H &  $<$0.07        & $<$1.2          & 16.3  & $>$0.005 & $<$4.3 \\
 & \\
 6I &  $<$0.08        & $<$1.5          & 16.7  & $>$0.01  & $<$4.4\\
 & \\
 7I &  $<$0.025       & $<$0.44         & 16.75 & $>$0.04  & $<$3.9 \\
 & \\
 8I &  $<$0.014       & $<$0.25         & 16.65 & $>$0.06  & $<$3.7 \\
 & \\
 9I &  $<$0.006       & $<$0.11         & 16.9  & $>$0.2   & $<$3.3 \\
 & \\
 10I &  $<$0.03       & $<$0.55         & 17.1  & $>$0.07  & $<$4.0 \\
 & \\
 11I &[0.010$\pm$0.002]&[0.17$\pm$0.04] &[16.95]&[0.17]    &[3.5]\\
 & \\
 12I &[0.010$\pm$0.002]&[0.17$\pm$0.04] &[18.15]&[2.7]     &[3.5]\\
 & \\
 13I &[0.010$\pm$0.002]&[0.17$\pm$0.04] &[17.75]&[1.1]     &[3.5]\\
 & \\
\enddata
\tablenotetext{a}{Components (1--5) are HV; components (6--13) are IV.}
\tablenotetext{b}{Ratio assumed equal for components 11I--13I in fits to C II* profile.}
\tablenotetext{c}{Assuming $T$ = 9000$\pm$2000 K, from $b$-values for C II, N II, Mg II, Si II, and Fe II in component 2H.  Limits for components 6I--9I assume $N$(C II*) $<$ 5 $\times$ 10$^{10}$ cm$^{-2}$.}
\tablenotetext{d}{From $N$(C II) + $N$(C III), $N$(N II) + $N$(N III), $N$(S II) + $N$(S III), and/or $N$(Si II) + $N$(Si III) assuming solar reference abundances, adjusted for adopted total HV or IV log[$N$(H II)] (see Table 4).}
\end{deluxetable}

\clearpage

\begin{deluxetable}{lccccccc}
\tabletypesize{\scriptsize}
\tablecolumns{8}
\tablenum{6}
\tablecaption{HV and IV  Gas toward $\zeta$ Ori, 23 Ori, $\tau$ CMa\tablenotemark{a} \label{tab:hvgprop}}
\tablewidth{0pt}
\tablehead{
\multicolumn{4}{c}{ }&
\multicolumn{1}{c}{HV Models}&
\multicolumn{2}{c}{ }&
\multicolumn{1}{c}{IV Models}\\
\multicolumn{1}{c}{Property}&
\multicolumn{1}{c}{$\zeta$ Ori HV}&
\multicolumn{1}{c}{23 Ori HV\tablenotemark{b}}&
\multicolumn{1}{c}{$\tau$ CMa HV}&
\multicolumn{1}{c}{A/B/C\tablenotemark{c}}&
\multicolumn{1}{c}{$\zeta$ Ori IV}&
\multicolumn{1}{c}{23 Ori IV\tablenotemark{b}}&
\multicolumn{1}{c}{D/E/F\tablenotemark{d}}}
\startdata
log[$N$(H II)] (cm$^{-2}$) & 17.88$^{+0.05}_{-0.06}$ & 17.8$\pm$0.1 & 17.0$\pm$0.1 &
     \nodata & 18.44$\pm$0.05 & 17.85$\pm$0.15 & \nodata \\
$N$(H I)/$N$(H II) & $<$0.001 & $<$0.01 & $<$0.05 & \nodata & 0.001 & $<$0.02 & \nodata \\
 & \\
$N$(C III)/$N$(C II) & 1.7$\pm$0.8 & $<$0.4 & \nodata & 0.7/6.6/3.6 & 1.0$\pm$0.5 & \nodata & 0.12/0.23/4.9 \\
$N$(C IV)/$N$(C II) & $<$0.004 & $<$0.004 & $<$0.019 & 0.02/0.11/0.0005 & 0.011$\pm$0.002 & \nodata & 0.011/0.013/0.17 \\
$N$(N III)/$N$(N II) & 1.4$\pm$0.3 & $<$0.1 & \nodata & 1.0/2.0/0.5 & 2.6$\pm$0.7 & \nodata & 0.16/0.23/1.8 \\
$N$(Al III)/$N$(Al II) & 2.6$\pm$0.3 & 0.3$\pm$0.1 & $<$1.5 & \nodata & 2.2$\pm$0.4 & 1.0$\pm$0.7 & \nodata \\
$N$(Si III)/$N$(Si II) & 4.4$\pm$1.0 & 3.0$\pm$0.5 & 0.7$\pm$0.1 & 17/240/141 & 6.2$\pm$1.3 & 4.4$\pm$2.1 & 0.13/0.34/537 \\
$N$(Si IV)/$N$(Si II) & $<$0.04 & $<$0.06 & $<$0.16 & 2.2/34/10 & 0.13$\pm$0.01 & \nodata & 0.031/0.054/49 \\
$N$(S III)/$N$(S II) & 2.7$\pm$0.6 & $<$0.6 & \nodata & 1.6/42/20 & 6.2$\pm$1.1 & \nodata & 0.44/1.0/91 \\
$N$(Fe III)/$N$(Fe II) & 11.$\pm$4. & \nodata & \nodata  & 52/117/71 & \nodata & \nodata & \nodata \\
 & \\
D(C)  &   $-$0.1$\pm$0.1 & $>$$-$0.2      & (0.0)     & \nodata & 0.0$\pm$0.1 & \nodata & \nodata \\
D(Al) & $-$0.8$\pm$0.05 & $-$0.5$\pm$0.1 & $>$$-$0.4 & \nodata & $-$0.85$\pm$0.05 & $-$0.35$\pm$0.15 & \nodata \\
D(Si) & $-$0.4$\pm$0.1 & $-$0.1$\pm$0.1 & 0.0$\pm$0.05 & \nodata & $-$0.35$\pm$0.1 & $-$0.15$\pm$0.15 & \nodata \\
D(Fe) & $-$0.95$\pm$0.15 & \nodata & \nodata & \nodata & \nodata & \nodata & \nodata \\
 & \\
log[$N$(C II*)/$N$(C II)] & $-$2.10$^{+0.08}_{-0.09}$ & $-$1.53$^{+0.05}_{-0.04}$ & $<$$-$1.95 & \nodata & $-$2.04$^{+0.04}_{-0.06}$ & $-$1.16$^{+0.11}_{-0.16}$ & \nodata \\
$<n_e>$ (cm$^{-3}$)\tablenotemark{e} & 0.14$^{+0.05}_{-0.04}$ & 0.50$^{+0.08}_{-0.08}$ & $<$0.18 & \nodata & 0.16$^{+0.04}_{-0.04}$ & 1.15$^{+0.44}_{-0.40}$ & \nodata \\
log[2$<n_{\rm H}>$$T$] (cm$^{-3}$K) & 3.4$\pm$0.2 & 3.9$\pm$0.1 & $<$3.5 & \nodata & 3.45$\pm$0.2 & 4.25$\pm$0.2 & \nodata \\
$N$(H II)/$<n_{\rm H}>$ (pc) & 1.8 & 0.4 & $>$0.2 & \nodata & 5.1 & 0.2 & \nodata \\
\enddata
\tablenotetext{a}{Total or mean values over all high- or intermediate-velocity components.  Uncertainties are 1-$\sigma$; limits are 2-$\sigma$.  Depletions are relative to reference abundances of Anders \& Grevesse 1989 and Grevesse \& Noels 1993.}
\tablenotetext{b}{Welty et al. 1999.}
\tablenotetext{c}{Shock travelling toward us at 95--100 km s$^{-1}$ ($\S$5.3.1; Table 7).}
\tablenotetext{d}{Shock travelling away from us, produced by impact of 95--100 km s$^{-1}$ flow upon low-velocity material ($\S$5.3.2; Table 7).}
\tablenotetext{e}{Assumes $T$ = 9000$\pm$2000 K for $\zeta$ Ori, $T$ = 8000$\pm$1000 K for 23 Ori and $\tau$ CMa.}
\end{deluxetable}

\clearpage
 
\begin{deluxetable}{lcccc}
\tabletypesize{\scriptsize}
\tablecolumns{5}
\tablenum{7}
\tablecaption{$\zeta$ Ori --- HV Ion Ratios\tablenotemark{a} \label{tab:ratios}}
\tablewidth{0pt}
\tablehead{
\multicolumn{1}{c}{Element}&
\multicolumn{1}{c}{(X/S)$_{\odot}$}&
\multicolumn{1}{c}{[X II/S II]}&
\multicolumn{1}{c}{[X III/S III]}&
\multicolumn{1}{c}{[X/S]}}
\startdata
C  &    1.28 &   +0.13$^{+0.09}_{-0.11}$ & $-$0.09$^{+0.17}_{-0.28}$ & 
     $-$0.02$^{+0.12}_{-0.16}$ \\
 & \\
N  &    0.70 &   +0.38$^{+0.09}_{-0.11}$ &   +0.09$^{+0.09}_{-0.10}$ & 
       +0.19$^{+0.05}_{-0.07}$ \\
 & \\
Al & $-$0.79 & $-$0.70$^{+0.09}_{-0.11}$ & $-$0.72$^{+0.05}_{-0.05}$ & 
     $-$0.72$^{+0.05}_{-0.05}$ \\
 & \\
Si &    0.28 & $-$0.44$^{+0.09}_{-0.09}$ & $-$0.23$^{+0.09}_{-0.12}$ & 
     $-$0.28$^{+0.08}_{-0.10}$ \\
 & \\
Fe &    0.24 & $-$1.37$^{+0.11}_{-0.14}$ & $-$0.77$^{+0.13}_{-0.18}$ & 
     $-$0.87$^{+0.13}_{-0.16}$ \\
 & \\  
\enddata
\tablenotetext{a}{Uncertainties are 1-$\sigma$; columns 3--5 are relative to solar ratios given in column 2 (Anders \& Grevesse 1989; Grevesse \& Noels 1993).}
\end{deluxetable}

\clearpage

\begin{deluxetable}{cccccc}
\tabletypesize{\small}
\tablecolumns{6}
\tablenum{8}
\tablecaption{Shock Model Parameters \label{tab:mod1}}
\tablewidth{0pt}
\tablehead{
\multicolumn{1}{c}{Model} &  
\multicolumn{1}{c}{$v_s$} & 
\multicolumn{1}{c}{$v_0$} & 
\multicolumn{1}{c}{$n_0$} & 
\multicolumn{1}{c}{$B_0$} & 
\multicolumn{1}{c}{W$_{\rm rad}$\tablenotemark{a}}\\
\multicolumn{1}{c}{ }&
\multicolumn{1}{c}{(km s$^{-1}$)}&
\multicolumn{1}{c}{(km s$^{-1}$)}&
\multicolumn{1}{c}{(cm$^{-3}$)}&
\multicolumn{1}{c}{($\mu$G)}&
\multicolumn{1}{c}{(erg cm$^{-2}$ s$^{-1}$ \AA$^{-1}$)}}
\startdata
\multicolumn{6}{c}{High-Velocity Gas} \\
\tableline
A  &   95    &  -12       &   0.0032   & 0.5 & $1 \times 10^{-7}$    \\
B  &   95    &  -14       &   0.0032   & 0.5 & $1 \times 10^{-5}$    \\
C  &  100    &  -38       &   0.01     & 2.5 & $1 \times 10^{-5}$    \\
\cutinhead{Intermediate-Velocity Gas}
D  &   95    &  -14       &   0.010    & 0.2 & 0                      \\
E  &  100    &   -7       &   0.020    & 1.1 & 0                      \\
F  &  100    &   -7       &   0.020    & 1.1 & $5.1 \times 10^{-5}$    \\
\enddata
\tablenotetext{a}{For diluted Kurucz (1979) model spectrum, near 1000 \AA.}
\end{deluxetable}

\clearpage
 
\begin{deluxetable}{lcccccc}
\tabletypesize{\scriptsize}
\tablecolumns{7}
\tablenum{9}
\tablecaption{Depletions and Dust Destruction \label{tab:depori}}
\tablewidth{0pt}
\tablehead{
\multicolumn{1}{c}{Element}&
\multicolumn{1}{c}{A$_{\odot}$\tablenotemark{a}}&
\multicolumn{1}{c}{Galactic\tablenotemark{b}}&
\multicolumn{1}{c}{23 Ori\tablenotemark{b}}&
\multicolumn{1}{c}{$\zeta$ Ori\tablenotemark{c}}&
\multicolumn{1}{c}{\% Destroyed\tablenotemark{d}}\\
\multicolumn{1}{c}{ }&
\multicolumn{1}{c}{ }&
\multicolumn{1}{c}{Warm/Cold}&
\multicolumn{1}{c}{HV/WLV/SLV}&
\multicolumn{1}{c}{HV/A/DE}}
\startdata
C  &8.55& $-$0.4/$-$0.4 & $>-$0.2/$-$0.31/$-$0.31 & $-$0.12/\nodata/\nodata&
     60--67\\
 & \\
N  &7.97& $-$0.1/$-$0.1 & \nodata/$-$0.15/$-$0.15 & +0.08/\nodata/\nodata& 
     \nodata\\
 & \\
Al &6.48& $-$1.1/$-$2.4 &  $-$0.5/$-$1.36/$-$1.71 & $-$0.82/$-$1.43/$-$1.84 &
      9--13\\
 & \\
Si &7.55& $-$0.4/$-$1.3 &  $-$0.1/$-$0.52/$-$0.89 & $-$0.39/$-$0.59/$-$0.72 &
      0--20\\
 & \\
Fe &7.51& $-$1.4/$-$2.2 & \nodata/$-$1.49/$-$1.92 & $-$0.97/$-$1.46/$-$1.85 &
      6--8\\
 & \\
\enddata
\tablenotetext{a}{Logarithmic solar abundances from Anders \& Grevesse 1989 and Grevesse \& Noels 1993.  H = 12.0.}
\tablenotetext{b}{Welty et al. 1999; WLV = ``weak, low-velocity'' components with log[$N$(H)] = 19.6 cm$^{-2}$; SLV = ``strong, low-velocity'' components with log[$N$(H)] = 20.7 cm$^{-2}$.}
\tablenotetext{c}{Welty et al., in prep.; A = components near 0 km s$^{-1}$; DE = strong components near +25 km s$^{-1}$.}
\tablenotetext{d}{Derived from $\zeta$ Ori HV (post-shock) versus $\zeta$ Ori A or Galactic warm cloud (assumed pre-shock) values.}
\end{deluxetable}

\clearpage

\begin{deluxetable}{llcrrcr}
\tabletypesize{\scriptsize}
\tablecolumns{7}
\tablenum{10}
\tablecaption{$\tau$ CMa --- HV Gas Equivalent Widths and Column Densities \label{tab:tcma}}
\tablewidth{0pt}

\tablehead{
\multicolumn{1}{c}{Ion}&
\multicolumn{1}{c}{$\lambda_{\rm vac}$\tablenotemark{a}}&
\multicolumn{1}{c}{ }&
\multicolumn{1}{c}{log(f$\lambda$)\tablenotemark{a}}&
\multicolumn{1}{c}{W$_{\lambda}$\tablenotemark{b}}&
\multicolumn{1}{c}{ }&
\multicolumn{1}{c}{log($N$)}\\
\multicolumn{1}{c}{ }&
\multicolumn{1}{c}{(\AA)}&
\multicolumn{2}{c}{ }&
\multicolumn{1}{c}{(m\AA)}&
\multicolumn{1}{c}{ }&
\multicolumn{1}{c}{(cm$^{-2}$)}}
\startdata
C I   &1560.3092 &G& 2.050 & $<$1.9\tablenotemark{c}       & & $<$12.09 \\
C II  &1334.5323 &E& 2.229 & 53.6$\pm$1.2                  & & 13.55$^{+0.01}_{-0.02}$ \\
C II* &1335.6627 &E& 1.229 & \multicolumn{2}{c}{blend with C II*}& $<$11.60 \\
      &1335.7077 &E& 2.183 & $<$0.8                        &B& \nodata \\
C IV  &1548.195  &G& 2.470 & $<$2.8\tablenotemark{c}       & & $<$11.84 \\
      &1550.770  &G& 2.169 & $<$2.8\tablenotemark{c}       & &\nodata\\
 &\\
N V   &1238.821  &G& 2.286 & $<$3.7                        & & $<$12.23 \\
      &1242.804  &G& 1.986 & $<$3.5                        & &\nodata\\
 &\\
O I   &1302.1685 &G& 1.831 & $<$3.2\tablenotemark{c}       & & $<$12.61 \\
 &\\
Mg I  &2026.4768 &E& 2.369 & $<$2.7\tablenotemark{c}       & & $<$11.81 \\
Mg II &2796.352  &E& 3.234 & 58.6$\pm$4.6\tablenotemark{c} & & 12.22$^{+0.04}_{-0.04}$ \\
      &2803.531  &E& 2.933 & 25.1$\pm$4.4\tablenotemark{c} & & \nodata \\
 &\\
Al I  &2367.7762 &E& 2.398 & $<$1.8\tablenotemark{c}       & & $<$11.54 \\
Al II &1670.7874 &E& 3.486 & 5.9$\pm$2.4                   & & 11.11$^{+0.15}_{-0.22}$ \\
Al III&1854.7164 &E& 3.017 & $<$3.4\tablenotemark{c}       & & $<$11.30 \\
      &1862.7895 &E& 2.716 & $<$3.4\tablenotemark{c}       & & \nodata \\
 &\\
Si I  &2515.0725 &E& 2.723 & $<$2.6\tablenotemark{c}       & & $<$11.34 \\
Si II &1190.4158 &E& 2.474 & 5.5$\pm$1.7                   & & 12.30$^{+0.01}_{-0.02}$ \\
      &1193.2897 &E& 2.775 & 10.9$\pm$1.7                  & & \nodata \\
      &1260.4221 &E& 3.104 & 23.3$\pm$1.5                  & & \nodata \\
Si II*&1264.7377 &G& 3.058 & $<$2.4                        & & $<$11.28 \\
Si III&1206.500  &E& 3.304 & 24.9$\pm$2.3                  & & 12.12$^{+0.03}_{-0.03}$ \\
Si IV &1393.755  &G& 2.855 & $<$2.7\tablenotemark{c}       & & $<$11.49 \\
      &1402.770  &E& 2.554 & $<$2.7\tablenotemark{c}       & & \nodata \\
&\\
S I   &1807.3113 &E& 2.239 & $<$1.7\tablenotemark{c}       & & $<$11.79 \\
S II  &1259.519  &E& 1.311 & $<$0.9                        & & $<$12.60 \\
S III &1190.208  &E& 1.421 & $<$1.4                        & & $<$12.70 \\
&\\
Fe I  &2484.0211 &E& 3.141 & $<$2.0\tablenotemark{c}       & & $<$10.82 \\
Fe II &2374.4612 &E& 1.871 & $<$2.7\tablenotemark{c}       & & $<$12.24 \\
 & \\
\enddata
\tablenotetext{a}{Vacuum wavelengths and $f$-values from Morton (1991), updated by Welty et al. 1999.  E = GHRS echelle; G = G160M.}
\tablenotetext{b}{B = major contributor to blend.  Uncertainties and limits are 2-$\sigma$.}
\tablenotetext{c}{Values from Trapero et al. 1996.}
\end{deluxetable}

\clearpage

\begin{deluxetable}{cccccccccccc}
\tabletypesize{\scriptsize}
\tablecolumns{12}
\tablenum{11}
\tablecaption{$\tau$ CMa --- HV Component Structure\tablenotemark{a} \label{tab:tcomp}}
\tablewidth{0pt}
\tablehead{
\multicolumn{1}{c}{$v_{\odot}$}&
\multicolumn{1}{c}{C II}&
\multicolumn{1}{c}{Mg II}&
\multicolumn{1}{c}{Si II}&
\multicolumn{1}{c}{Si III}\\
\multicolumn{1}{c}{(km s$^{-1}$)}&
\multicolumn{1}{c}{(10$^{12}$)}&
\multicolumn{1}{c}{(10$^{11}$)}&
\multicolumn{1}{c}{(10$^{11}$)}&
\multicolumn{1}{c}{(10$^{11}$)}}
\startdata
$-$96.4$\pm$0.1 & 12.6$\pm$1.1 & 7.0$\pm$0.3 & 9.5$\pm$0.4 & 1.3$\pm$0.3 \\
 1H             &  3.3$\pm$0.2 & 2.3$\pm$0.2 & 2.4$\pm$0.2 & 1.8$\pm$0.9 \\
 & \\
$-$87.8$\pm$0.1 & 17.3$\pm$1.6 & 7.3$\pm$0.4 & 8.5$\pm$0.6 & 8.4$\pm$0.6 \\
2H              &  5.2$\pm$0.6 & 3.8$\pm$0.3 & 4.1$\pm$0.4 & 5.0$\pm$0.6 \\
 & \\
$-$80.9$\pm$0.4 &  1.9$\pm$0.7 & 0.9$\pm$0.2 & (0.2) & 1.2$\pm$0.4 \\
3H              & (3.1)        & (2.1)       & (2.0) & (2.5) \\
 & \\
$-$69.7$\pm$0.2 &  4.2$\pm$0.2 & 1.5$\pm$0.2 & 1.8$\pm$0.2 & 2.4$\pm$0.3 \\
4H              &  4.9$\pm$0.4 & 2.9$\pm$0.5 & 2.4$\pm$0.5 & 4.3$\pm$0.7 \\
 & \\
\enddata
\tablenotetext{a}{Entries are N (cm$^{-2}$) and b (km s$^{-1}$).   Values in parentheses were fixed in the fitting.  Uncertainties are 1-$\sigma$.}
\end{deluxetable}

\end{document}